# Luminous mid-IR-selected type 2 quasars at cosmic noon in SDSS Stripe 82 – I. Selection, composite photometry, and spectral energy distributions

Ben Wang ●,[1,2]★ Joseph F. Hennawi ●,[1,3] Zheng Cai,[2] Gordon T. Richards,[4] Jan-Torge Schindler,[5] Nadia L. Zakamska ●,[6,7] Yuzo Ishikawa,[6] Hollis B. Akins[8] and Zechang Sun[2]

[1]*Leiden Observatory, Leiden University, NL-2333 CA Leiden, the Netherlands*
[2]*Department of Astronomy, Tsinghua University, Beijing 100084, China*
[3]*Department of Physics, University of California Santa Barbara, Broida Hall, Santa Barbara, CA 93106-9530, USA*
[4]*Department of Physics, Drexel University, 32 S 32nd Street, Philadelphia, PA 19104, USA*
[5]*Hamburg Observatory, Gojenbergsweg 112, D-21029 Hamburg, Germany*
[6]*Department of Physics and Astronomy, Bloomberg Center, Johns Hopkins University, Baltimore, MD 21218, USA*
[7]*Institute for Advanced Study, Princeton University, Princeton, NJ 08544, USA*
[8]*Department of Astronomy, The University of Texas at Austin, 2515 Speedway Blvd, Stop C1400, Austin, TX 78712, USA*

## ABSTRACT

We analyse 23 spectroscopically confirmed type 2 quasars (QSOs) selected from *Wide-field Infrared Survey Explorer* 22 μm band in the Sloan Digital Sky Survey Stripe 82 region, focusing on their multiband photometry and spectral energy distributions (SEDs). The 24 candidates were selected to be infrared (IR) luminous (flux$_{W4}$ > 5 mJy), optically faint ($r$ > 23), or with red colour ($r - W4$ > 8.38). Gemini/Gemini Near-InfraRed Spectrograph and Keck/Low-Resolution Imaging Spectrometer observations confirm 23 to be type 2 QSOs at $z = 0.88$–3.49. Multiband photometry is used for SED fitting, covering 0.1–10 μm in the rest frame. The IR emission is dominated by the dust torus, with an average luminosity of $L_{torus} = 10^{46.84}$ erg s$^{-1}$. We present three possibilities for the origin of the rest-ultraviolet/optical: scattered light, stellar emission, and the reddened accretion disc. Assuming an obscured:unobscured ratio of 1:1, the targets have bolometric luminosities of $L_{bol} = 10^{46.28}$–$10^{48.08}$ erg s$^{-1}$ and supermassive black hole masses of $10^{8.18}$–$10^{9.98}$ M$_\odot$, averaging $L_{bol} = 10^{47.04}$ erg s$^{-1}$ and $M_{BH} = 10^{8.94}$ M$_\odot$, assuming the Eddington limit. Compared to previous type 2 active galactic nuclei SEDs, our targets have a brighter dust torus and redder optical–IR colour. By comparing the SED to *JWST* 'little red dots' (LRDs), we find that these IR-selected type 2 QSOs have similar SED shapes to the LRDs. This survey demonstrates mid-IR selection as an efficient method to find luminous type 2 QSOs and the composite photometry generated by this sample provides a guide for finding more type 2 QSOs at higher redshift in the future.

## 1 INTRODUCTION

Understanding how supermassive black holes (SMBHs) and their host galaxies coevolve in the Universe is an unsolved problem. The classic Soltan argument (Soltan 1982) was raised to understand the build-up of SMBHs. In this picture, the integrated quasar (QSO) luminosity density is related to the mass density of relic black holes. The contributions from obscured QSOs are included in defining the QSO luminosity density. In a classical 'unified' model (Antonucci 1993; Urry & Padovani 1995) for active galactic nuclei (AGNs), observers can detect the objects with both continuum emission from the accretion disc region and broad emission lines produced deep within the potential well of the host black hole. These objects are called type 1 AGNs or unobscured AGNs. If the line of sight to the broad emission line and the central continuum regions are obscured by the dusty region, observers can only detect narrow emission lines. Then, these objects are called type 2 AGNs or obscured AGNs. Another scenario is that the difference between type 1 and type 2 QSOs is due to evolutionary effects. In this picture, type 1 and type 2 QSOs are at different evolution phases of SMBH/galaxy coevolution, especially after the galaxy merger (e.g. Sanders et al. 1988; Hopkins et al. 2006). Constraining the type 1:type 2 ratio at all cosmic times can help to better understand the SMBH accretion growth, and further reveal the physical processes producing these two different kinds of QSOs.

Over the past decades, type 1 QSOs have been discovered in large numbers (more than 750 000 QSOs) by dedicated projects like Sloan Digital Sky Survey (SDSS; e.g. Schneider et al. 2007; Lyke et al. 2020) and the ongoing Dark Energy Spectroscopic Instrument

★ E-mail: bwang@mail.strw.leidenuniv.nl

(DESI; e.g. Chaussidon et al. 2023; DESI Collaboration 2024) survey. Luminous type 1 QSOs are also identified at high redshift $z > 7$ (e.g. Bañados et al. 2018; Wang et al. 2021). For type 2 QSOs, sizable samples of type 2 QSOs have been identified from mid-infrared (IR), X-ray, and the SDSS before *JWST* (e.g. Zakamska et al. 2003; Haas et al. 2004; Alonso-Herrero et al. 2006; Martínez-Sansigre et al. 2006; Brand et al. 2007; Alexandroff et al. 2013; Lacy et al. 2013; Yuan, Strauss & Zakamska 2016). However, the majority of these type 2 QSOs are low-luminosity objects at $z < 1$. The number of spectroscopically confirmed type 2 AGN at $z > 2$ is very limited (Lacy et al. 2013), leading to a poorly understood QSO luminosity function (QLF; Glikman et al. 2018).

Optical type 2 QSO surveys from SDSS spectroscopy (Zakamska et al. 2003, 2006; Reyes et al. 2008; Yuan et al. 2016) have produced the largest type 2 samples at low $z < 1$. Some blind narrow-band surveys also identified a small number of type 2 QSOs through the optical-continuum dark diffuse Ly $\alpha$ emission (e.g. Cai et al. 2017; Zhang et al. 2023; Li et al. 2024). But these surveys only cover small sky areas ($\sim$12 deg$^2$) and such selection methods are hard to apply to the whole sky. Dust obscuration makes type 2 QSOs faint in optical, so surveys in other wavelength bands are necessary to reveal the population of these obscured QSOs.

X-ray surveys provide another method to identify type 2 AGN (Polletta et al. 2006; Brandt & Alexander 2015; Peca et al. 2023, 2024) and purport to measure the type 2 QLF out to $z \sim 5$ (e.g. Hasinger 2008; Ueda et al. 2014), but there are important caveats: (1) at high-$z$ they rely almost entirely on photo-$z$s whose fidelity for type 2s has not been demonstrated; (2) they cover only $\lesssim$10 deg$^2$ areas, insufficient to identify significant numbers of luminous high-$z$ type 2s; (3) as they lack spectroscopy, X-ray hardness ratios are used distinguish type 1s from type 2s, but it is unclear how this correlates the canonical classifications based on emission line width (e.g. Zakamska et al. 2003); and (4) Compton thick objects with high photoelectric obscuring columns $N_H > 10^{24}$ cm$^{-2}$ will be absent from X-ray samples.

In contrast, mid-IR selection provides an unbiased and more complete way to find obscured QSOs. According to the current knowledge of type 2 QSO spectral energy distributions (SEDs), both type 1 and type 2 AGN SEDs peak in the mid-IR due to hot dust reprocessing the optical/ultraviolet (UV) radiation from the accretion disc. Exploiting this, colour-selection wedges based on mid-IR photometry from *Spitzer* (Lacy et al. 2004; Stern et al. 2005; Polletta et al. 2008; Donley et al. 2012) and later *Wide-field Infrared Survey Explorer* (*WISE*; Assef et al. 2013; Glikman et al. 2018) were used to select AGN candidates, later confirmed via optical and IR spectroscopy (Lacy et al. 2013; Glikman et al. 2018). However, the redshift distribution of mid-IR-selected type 2s exhibits a precipitous drop at $z > 1$ (see fig. 7 in Lacy et al. 2013 or fig. 10 in Glikman et al. 2018) resulting in only 18 spectroscopic confirmed luminous $z > 2$ type 2s. Another class of *WISE*-selected AGNs is '*W1W2* dropouts' or hot dust-obscured galaxies (Hot DOGs), which are required to have very red mid-IR SEDs. Published spectra of ~60 Hot DOGs indicate that ~40 are type 2s at $z \gtrsim 2$ (Wu et al. 2012; Tsai et al. 2015), but the primary difference is that they are nearly an order of magnitude less abundant on the sky than DOGs (252 Hot DOGs in 32 000 deg$^2$; Eisenhardt et al. 2012; Assef et al. 2015), implying that they likely constitute the (mid-IR) reddest subset of the type 2 population. These *WISE*-selected QSOs are still mainly at $z < 2$ and are not comparable to the type 1 QSO number density. Some extremely red quasars (ERQs) and heavily reddened quasars (HRQs) are also selected using the IR selection (e.g. Banerji et al. 2015; Ross et al. 2015; Zakamska et al. 2016; Hamann et al. 2017; Alexandroff et al. 2018; Temple et al. 2019; Zakamska & Alexandroff 2023). However, these objects are selected with an optical magnitude cut like $i < 20.5$ which will have a luminosity bias (Banerji et al. 2015).

These numerous multiwavelength type 2 QSO surveys provide a large sample to reveal the ratio between type 2 and type 1 (type 2:type 1) at low redshift. According to the unification models, the number density of type 1 and type 2 QSOs should be comparable through cosmic time, both peaking at $z \sim 2.5$ (Richards et al. 2006; Ross et al. 2013). Decades of AGN/QSO censuses across the electromagnetic spectrum (e.g. Zakamska et al. 2003; Assef et al. 2013; Lacy et al. 2013; Yan et al. 2013; Brandt & Alexander 2015) have led to the consensus that the type 2:type 1 ratio is comparable at low redshift ($z < 1$)(e.g. Reyes et al. 2008; Lawrence & Elvis 2010). Merloni et al. (2014) claim that the obscured fraction is about 50 per cent based on the X-ray versus optical SEDs of 1310 AGNs at $0 < z < 3$.

However, the putative *JWST* discovery of a large population of obscured high-$z$ ($4 \lesssim z \lesssim 9$) AGN would indicate a higher obscured fraction (e.g. Kocevski et al. 2023; Greene et al. 2024; Matthee et al. 2024). The new *JWST* AGNs appear to suggest that type 2:type 1 ratios are more like $\sim 10^{3-4}$:1. Following the Soltan argument, there is as much SMBH growth at $z \gtrsim 4$ as at later times. But the redshift distribution of these 'little red dots' (LRDs) is mainly at $z = 4-9$ (e.g. Kocevski et al. 2023; Akins et al. 2024). It is still unclear if a large obscured population has been missed at $2 < z < 4$, leading to large uncertainty in tracing SMBH accretion growth.

Moreover, the physical properties of these high number density LRDs discovered by *JWST* remain unknown. Whether they are AGN dominated or galaxy light dominated is hotly debated (e.g. Harikane et al. 2023; Kocevski et al. 2023; Akins et al. 2024; Labbé et al. 2025). Some investigations argue that they could be obscured AGN in blue galaxies, or unobscured AGN in red galaxies (Kocevski et al. 2023). Some authors use obscured AGN with scattered light to explain the 'V-shape' of the SEDs (Greene et al. 2024; Labbé et al. 2025). Matthee et al. (2024) claim they could be in some transition phases. Akins et al. (2024) make extreme assumptions to see if they could be pure luminous AGN or star-forming galaxies. Their faintness and relatively low bolometric luminosity make it difficult to uncover their true nature. The new *JWST* AGNs are all at high-$z$ ($4 \lesssim z \lesssim 9$) with $L_{bol} \sim 10^{45-46}$ erg s$^{-1}$ and are about an order of magnitude fainter than the typical $L_{bol,*} \sim 10^{46.5}$ erg s$^{-1}$ QSOs at $z \sim 2$ that are believed to dominate the QSO luminosity density (e.g. Fontanot et al. 2007; Kulkarni, Worseck & Hennawi 2019). Such a large population of *JWST* LRDs would indicate that the obscured fraction is much higher than we expect, especially for the faint AGNs.

Investigations of low-redshift QSOs have shown that more luminous QSOs are less likely to be obscured (La Franca et al. 2005; Simpson 2005; Maiolino et al. 2007; Burlon et al. 2011). However, we do not see much evidence in Lusso et al. (2015), although it was at higher $z$. This trend that more luminous QSOs are less likely to be obscured could be explained by the inner torus structure receding outwards, which will reduce the dust covering factor and make the central engine observable with increasing luminosity (Lawrence 1991; Toba et al. 2014). Another possible scenario is that AGN blows out the obscuring material after exceeding the Eddington ratio so that the obscured fraction is dominated by the radiation pressure (Walter et al. 2009). Finding more luminous type 2 QSOs at high redshift can provide the observational evidence to test these scenarios, and may help to explain the lower numbers of bright obscured AGNs.

Finding luminous type 2 QSOs relies highly on our understanding of their SED (e.g. Hickox et al. 2017; Zakamska et al. 2019). Mid-IR selection has been proven to be an efficient way to uncover the type 2 population because it probes the peak of the SED. However, the current composite SED is mainly based on the type 2 sample at low redshift $z < 1$ (e.g. Hickox et al. 2017). We need to generate a new SED model of high-$z$ type 2s to see if there is a redshift evolution, and use it to guide us to search for type 2s at higher redshift. Polletta et al. (2007) collected X-ray-selected AGN samples and found obscured AGNs have hard X-ray spectra, consistent with being absorbed. Lusso et al. (2013) use the SED fitting to reveal the obscured fraction and the result favours a torus optically thin to mid-IR radiation. The SED fitting on type 2 QSOs in Hickox et al. (2017) shows type 2 QSOs have remarkably similar SEDs to type 1s. They conclude that the obscured QSOs can be efficiently selected from optical–IR colours. Fan et al. (2016) summarized the SED properties of Hot DOGs: the hot dust torus emission dominates the IR energy output but the cold dust emission is non-negligible.

Besides, most current LRD samples are photometrically selected and lack complete spectroscopic observation (e.g. Durodola, Pacucci & Hickox 2024; Pérez-González et al. 2024). Therefore, discussions of their AGN or galaxy components rely on photometry and SED fitting. One key to understanding the true nature of LRDs is to reveal if they have AGN hot dust, and this can be determined based on their SEDs. Akins et al. (2024) fit galaxy and AGN SED model on 434 LRDs selected from the 0.54 deg$^2$ COSMOS(Cosmological Evolution Survey)-Web survey. The SED fitting results in a cold dust temperature of about 200 K. Based on the same LRD sample, Casey et al. (2024) argue the SEDs are thought to peak at $\sim$100 K (rest-frame 20–30 μm) regardless of AGN or star formation dominating. Comparing the SED of LRDs and the $z \sim 2$ type 2 QSOs can help us to understand if LRDs can have hot dust and figure out if LRDs are related to obscured QSOs.

In this project, we conduct a pilot type 2 QSO survey in the SDSS Stripe 82 region using a *WISE*-based IR selection (Ishikawa et al. 2023). We want to find the most luminous type 2 QSOs at $z > 2$. We work at the bright end because (1) the bright targets are more easily confirmed with spectroscopy even with ground-based telescopes; (2) bright objects are free from concerns about disentangling stellar and nuclear emission that have plagued studies of fainter AGN with *JWST* (e.g. Greene et al. 2024). Previous investigations claim that the more luminous QSOs are less likely to be obscured; we can test this scenario at high redshift. Based on the type 2 QSO SED, we designed a red colour selection ($r - W4 > 8.38$) to find 24 candidates. Spectroscopic follow-up observations using the Gemini North and Keck telescopes were conducted to confirm these candidates. We plan to publish our results in two papers. In this paper, we present the first part of our results: the composite photometry and the SED fitting results. The spectroscopic results will be discussed in the next paper. This paper is divided as follows. In Section 2, we introduce the target selection and observation and shortly go through the diversity of the spectra. In Section 3, we conduct the SED fitting, generate the composite photometry, and compare the SED model to previous works. In Section 4, we make a comparison to *JWST* LRDs.

## 2 TARGET SELECTION AND OBSERVATION

In this section, we describe our mid-IR selection for the type 2 candidates and the spectroscopic observations. The observation results are also discussed briefly.

### 2.1 Target selection

In order to identify and investigate the elusive high-$z$ luminous type 2 QSO population, 24 bright type 2 QSO candidates in SDSS Stripe 82

are photometrically selected from the *WISE* 22 μm (*W4*) observation (Ishikawa et al. 2023). The selection will be discussed in detail in a future paper. Here we just describe it briefly. All the magnitudes in this paper are AB magnitude (ABmag). Since QSOs have a hot-dust peak in the SEDs at 10 μm, at $z > 2$, they will peak in the *WISE W4* band (22 μm). To select IR-luminous type 2 QSOs, we require our candidates to have a signal-to-noise ratio (SNR) above 5 in *W4* band with $12.62 < W4 < 14.62$ ABmag ($f_{W4} > 5$ mJy). If this *W4* flux is powered by the reprocessing of the AGN big blue bump (BBB) from the dust torus, then the bolometric luminosity will be around $10^{47}$ erg s$^{-1}$ according to the current knowledge of the type 2 QSO SED. This luminosity corresponds to $M_{1450} \simeq -28$ at $z \sim 2$, which is at the bright end of the QSO luminosity function. The lower limit of the magnitude range is to exclude bright stars or bright galaxies at low redshift. We further require candidates to be either non-detected in SDSS ($r > 23$ in the SDSS Stripe 82 catalogue) or with a red colour cut: $r - W4 > 8.38$. Under this selection, the ratio of type 2 and type 1 QSOs are around $0.55{:}0.65$ deg$^{-2}$ in the SDSS Stripe 82 region at the bright end ($L_{bol} \sim 10^{47}$ erg s$^{-1}$).

Finally, the candidates are matched with the *Spitzer* Infrared Array Camera (IRAC) catalogue. The catalogue used to match combined: (1) The *Spitzer*/HETDEX Exploratory Large-Area (SHELA) survey catalogue (Papovich et al. 2016). This survey covers $\sim$24 deg$^2$ at 3.6 and 4.5 μm and falls within the footprint of the SDSS Stripe 82 region. The catalogues reach limiting sensitivities of 1.1 mJy at both 3.6 and 4.5 μm. (2) The *Spitzer* IRAC Equatorial Survey (SpIES) survey catalogue (Timlin et al. 2016). This is a large-area survey of 115 deg$^2$ in the Equatorial SDSS Stripe 82 field. SpIES achieves 5-$\sigma$ depths of 6.13 mJy (21.93 AB) and 5.75 mJy (22.0 AB) at 3.6 and 4.5 μm, respectively. All the targets are matched to the combined catalogue to get the precise coordinates. Considering all the selection criteria, 24 candidates are identified from a total area of 164 deg$^2$. An example candidate is shown in Fig. 1. We summarize the selection to be

$$12.62 < W4 < 14.62 \text{ AB mag},$$
$$\text{SNR}_{W4} > 5, \tag{1}$$
$$\text{and } (r > 23 \text{ AB mag  or  } r - W4 > 8.38 \text{ AB mag}).$$

### 2.2 Observation and data reduction

To confirm the redshift of these candidates, we use Gemini Near-InfraRed Spectrograph (GNIRS; Elias et al. 2006) on the Gemini North telescope to conduct the spectroscopic observation. The candidates were observed under GN-2017B-Q-51 (PI: Gordon Richards) for several nights in 2017 September, 2017 October, and 2018 January. The grating was 32/mmSB which covers 0.9–2.4 μm with spectral resolution of $R \sim 1100$. The exposure time for each candidate was 2400 s and the observations were performed using an ABBA sequence. The results were first published in Ishikawa et al. (2023).

Based on the GNRIS observation, we identified 17 type 2 QSO candidates, six reddened type 1 QSO candidates (broad >2000 km s$^{-1}$ emission lines and bright continuum with SNR > 1), and one target remains inconclusive. We choose 18 targets (the 17 type 2 QSO candidates and the one inconclusive target) to conduct a follow-up observation using Keck Low-Resolution Imaging Spectrometer (LRIS; Oke et al. 1995), to better confirm the redshifts and reveal the rest-UV spectra. The observations were taken on the night of 2022 September 27, and over the course of a 3-night run from 2022 October 26–28. We observed these targets using a 1.0 arcsec long slit, $2 \times 2$ binning, 560 dichroic, and 600/4000 grating for the blue band with resolution ∼4 Å covering 3040–5630 Å. The grating for the red band is: 400/8500, 1.0 arcsec long slit, resolution around 7 Å, $\Delta\lambda = 4762$ Å with the central wavelength 7980 Å. The average exposure time for each candidate is 1800 s.

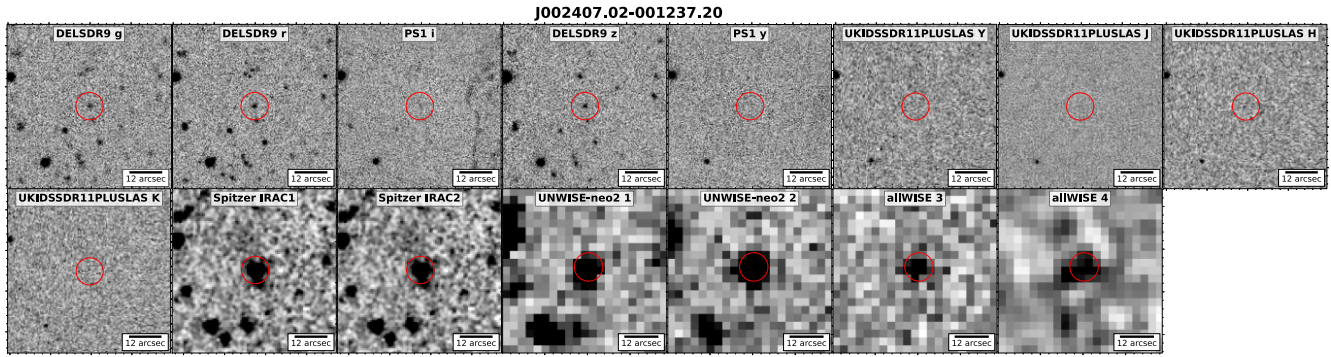

**Figure 1.** Cut-out imaging of one candidate. The images are from DESI Legacy Survey, PS1, UKIDSS, *Spitzer*, and *WISE*. The red central circle has a radius of 5 arcsec. This candidate has optical detections in the *g*, *r*, and *z* bands and has a red colour $r - W4 > 9.4$.

For the six reddened type 1 QSO candidates, we further observed two with the Keck Cosmic Web Imager Integral Field Spectrograph (KCWI; Morrissey et al. 2018), to better constrain their redshift. The observation was conducted on 2025 January 28. We used a large slicer, 2 × 2 binning, BL grating for the blue, and RL grating for the red. Each target was observed with 1800 s exposure.

The Gemini and Keck observation data were reduced using PYPEIT pipeline (Prochaska et al. 2020). This pipeline can automatically produce calibrated 2D spectra. We coadded the individual 2D spectra and used manual extraction to obtain the 1D spectra. The flux calibration and telluric correction were conducted using standard stars observed on the same night.

### 2.3 Emission lines and redshift estimates

For the GNIRS observation of 24 targets, we have identified emission lines (H α and [O III]) for 21 targets, and the redshifts are estimated according to these emission lines ($z_G$). The remaining three objects targeted with GNIRS have no detected emission lines with SNR > 3 in their spectra. To better confirm the redshift and study the rest-UV properties of these QSOs, we obtained Keck/LRIS or Keck/KCWI spectra for 20 targets, including 17 targets with redshifts $z_G$ from the GNIRS spectra, and three targets without $z_G$. We successfully identified Ly α emission lines in 13 targets and other emission lines (N V λ1240, C IV λ1549, and Mg II λ2798) in 11 targets. We estimate the redshift using these emission lines in LRIS or KCWI spectra ($z_K$).

We conducted two steps in estimating $z_K$. First, we manually check the spectra and identify the emission lines (like Ly α and C IV λ1549) to determine the rough spectroscopic redshifts. Then, we fit a Gaussian profile to every emission line to get accurate redshifts and line width.

For the 17 targets with $z_G$ estimates, the $z_K$ are consistent with $z_G$ with differences less than 0.01. We have also estimated the redshift for two targets without detectable lines in the GNIRS spectra because the bad weather conditions or the detectable lines fall into the telluric region. These two targets (J2239−0030 and J2239−0054) have Ly α and C IV λ1549 emissions with SNR > 10 in the LRIS spectra. So the spectroscopic redshift can be fitted. For another target without GNIRS spectroscopic redshift, J0047+0003, no emission lines or continuum are detected in the LRIS spectra either. This target still remains inconclusive. A deeper or wider wavelength coverage spectroscopic observation is needed to uncover the true nature of this target. The redshift distribution of the 23 identified targets is shown in Fig. 2.

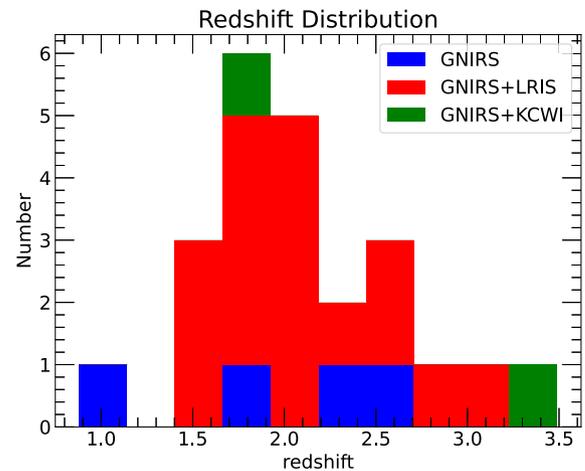

**Figure 2.** Redshift distribution of all the targets in the stacked plot. 12/23 identified targets have redshift >2. The targets with both GNIRS and LRIS spectra are plotted as red hatches, targets that have both GNIRS and KCWI spectra are plotted as green hatches, and the blue hatches are the targets only with GNIRS spectra. For these 23 identified targets, 17/23 are both GNIRS and LRIS confirmed, and 2/23 are both GNIRS and KCWI confirmed. The remaining four targets only have GNIRS spectra.

### 2.4 Efficiency of mid-IR selection

Among 24 targets, we identified 23 targets to be real type 2 or redden type 1 QSOs at $z \sim 2$, yielding a very high successful rate (∼96 per cent). Of the 23 confirmed targets, 12 are at $z > 2$ (∼52 per cent). Previous type 2 QSO survey using *Spitzer* like Lacy et al. (2013) finds 294 type 2 QSOs, but only 18 are at $z > 2$ (∼6 per cent). Our $r - W4$ selection shows a high successful rate and higher efficiency in isolating high $z > 2$ type 2 QSOs. This mid-IR selection is effective because it probes the peak of the AGN IR SED ($\lambda_{rest} \sim 10$ μm) by adopting the reddest possible mid-IR band, while excluding objects with AGN emission at bluer wavelengths by demanding a red $r - [24$ μm$]$ colour.

## 2.5 Spectral diversity

The identified 23 targets show different features on the near-IR photometry and the GNIRS spectra. Some targets have strong $K$-band detection in UKIDSS and continuum detection in the Gemini/GNIRS spectra, making them not typical type 1 QSOs but also not typical type 2 QSOs. Here we refer to them as 'reddened type 1 QSO'. The reddened type 1 QSO was defined in previous studies as having broad lines in rest-optical and a red continuum in rest-UV/optical (e.g. Glikman et al. 2018). Here we use a slightly different definition.[1] The density of redden type 1 QSO in our sample is less than 4.3 per cent (1/23).

Our sample shows a large diversity of spectra and some examples are shown in Figs 3, 4, and 5. Fig. 4 shows two 'LRD-like' spectra (broad H$\alpha$ emission line, narrow Ly$\alpha$ emission line, and no other emission lines, namely J0221+0050 and J2258+0022. Two type 2 QSO example spectra are shown in Fig. 3: J2229+0022 and J0112+0016, the latter of which is at the highest redshift end in our sample ($z = 2.99$). It has a narrow Ly$\alpha$ emission line and [O III] emission line. The target J2229+0022 is a type 2 QSO with strong narrow UV emission lines. The LRIS spectra show detection of Ly$\alpha$, C IV $\lambda$1549, and Mg II $\lambda$2798. The [O III] and H$\alpha$ lines are detected in GNIRS but with low SNR. The GNIRS spectrum of the only reddened type 1 QSO (J0213+0024) in our sample is shown in Fig. 5.

Interestingly, we find 3/23 targets have broad emission lines ($>4000 \text{ km s}^{-1}$). This is different from the typical type 2 QSOs with narrow emission lines ($<2000 \text{ km s}^{-1}$). Some of these targets show similar spectra to *JWST* broad-line AGNs and LRDs (see Fig. 4), making them possible low-$z$ counterparts of the *JWST* AGNs. A detailed discussion of the spectra will be presented in the next paper.

# 3 SPECTRA ENERGY DISTRIBUTIONS FITTING

In this section, we collect photometry from optical to mid-IR surveys for all 23 identified targets. To uncover the components and normalize all the photometry, we conduct the SEDs fitting using AGNFITTER. The SED fitting results are used to normalize the photometry and generate the new composite photometry of type 2 QSOs.

## 3.1 Aperture photometry

Given spectroscopic redshifts (from Keck/LRIS and Gemini/GNIRS) for 23 targets and photometric coverage from the optical to mid-IR in the SDSS Stripe 82 region, we can perform an analysis of the SEDs of our type 2 QSO sample.

Among the total 23 identified targets, 21/23 are detected in the DESI Legacy Survey, so we use the photometry in the DESI catalogue for these targets. We conduct aperture force photometry for the remaining two targets using a 2 arcsec aperture. All 23 targets are covered by the UKIRT Infrared Deep Sky Survey (UKIDSS; Lawrence et al. 2007) survey but only five targets are detected in $K$ band by matching to the catalogue. We performed a similar aperture forced photometry to obtain the photometry from the UKIDSS. We also apply aperture force photometry on The Pan-STARRS1 Surveys (PS1; Chambers et al. 2016) to get the flux in $i$, $y$ bands. For the *Spitzer* photometry, the targets are from two different surveys: SpIES and SHELA. The catalogue is matched to get the 3.6 and 4.5 µm flux. We also include the photometry in the $W1$ and $W2$ bands from the *UNWISE* survey and $W3$ and $W4$ bands from the *ALLWISE* catalogue. We use the flux and error ratio to calculate the SNR. If SNR > 2, we consider it as a detection. Otherwise, it is taken as a non-detection, and the magnitude will be shown in limit. We take the flux plus error to calculate the magnitude if the flux is negative. We summarize the photometry in AB magnitude and present it in Tables 1 and 2. A table with the flux of each target is shown in the appendix. The surveys and the bands used in this paper are summarized in Table 3.

## 3.2 Ly$\alpha$ forest correction

Since the Ly$\alpha$ forest of most targets falls in the $g$ band, we apply a Ly$\alpha$ forest correction using

$$\frac{F_{\text{true}}}{F_{\text{obs}}} = f_{\text{Ly}\alpha} \times \left(\frac{1.0}{F(z)}\right) + (1 - f_{\text{Ly}\alpha}), \tag{2}$$

where $f_{\text{Ly}\alpha}$ is the fraction of Ly$\alpha$ forest covers the photometry band. $F(z)$ is the mean Ly$\alpha$ transmitted flux at different redshift. We take the function from Becker et al. (2013) to get this value in different redshifts. Since most of our targets have redshifts from $z = 1.5$ to 3, the Ly$\alpha$ correction is tiny (around 0.04). We use this corrected flux for the SED fitting.

## 3.3 Fitting SED using AGNFITTER

To better conduct a normalization of all the photometry and understand what physical processes contribute to emission at each wavelength band, we use AGNFITTER to perform the SED fitting, combining all these available photometric data (see Section 3.1). AGNFITTER is a PYTHON algorithm implementing a Bayesian method to fit the SEDs of AGNs and galaxies (Calistro Rivera et al. 2016; Martínez-Ramírez et al. 2024).

The AGNFITTER fitting result is mainly used to normalize the photometry from different targets and reveal the existence of the hot dust torus. The AGNFITTER pipeline enables the characterization of four physical components including dust torus, black hole accretion disc, stellar emission, and cold dust.

*Dust torus.* This component is the key to understanding the IR properties of these type 2 QSOs, the model is from Stalevski et al. (2016). The templates were generated using the SKIRT radiative transfer code (Camps & Baes 2015, 2020), taking into account the geometry of a flared disc that is truncated at the dust sublimation region. The dust temperature in this model is set to 1250 K. The free parameters of this template are the torus inclination angle, incl, and the normalization parameter, TO.

*Accretion disc.* The model is based on THB21 (Temple et al. 2021). This black hole accretion disc model is based on a set of empirically derived composite SEDs of luminous QSOs. The free parameters for this mode are BBB reddening, $E(B-V)_{\text{bbb}}$, and the normalization parameter BB.

*Stellar emission.* We use the template from Bruzual & Charlot (2003), with metallicities for this component. This model is defined for stellar population ages ranging from $10^7$ to $10^{10}$ yr and stellar metallicities between 0.004 and 0.04. The parameter space for this model is: $\tau$ [time-scale of the exponential star formation history (SFH) in log year], age (galaxy age in the unit of log year), Z (metallicity), E(B–V)_gal (galaxy reddening), and the normalization parameter GA.

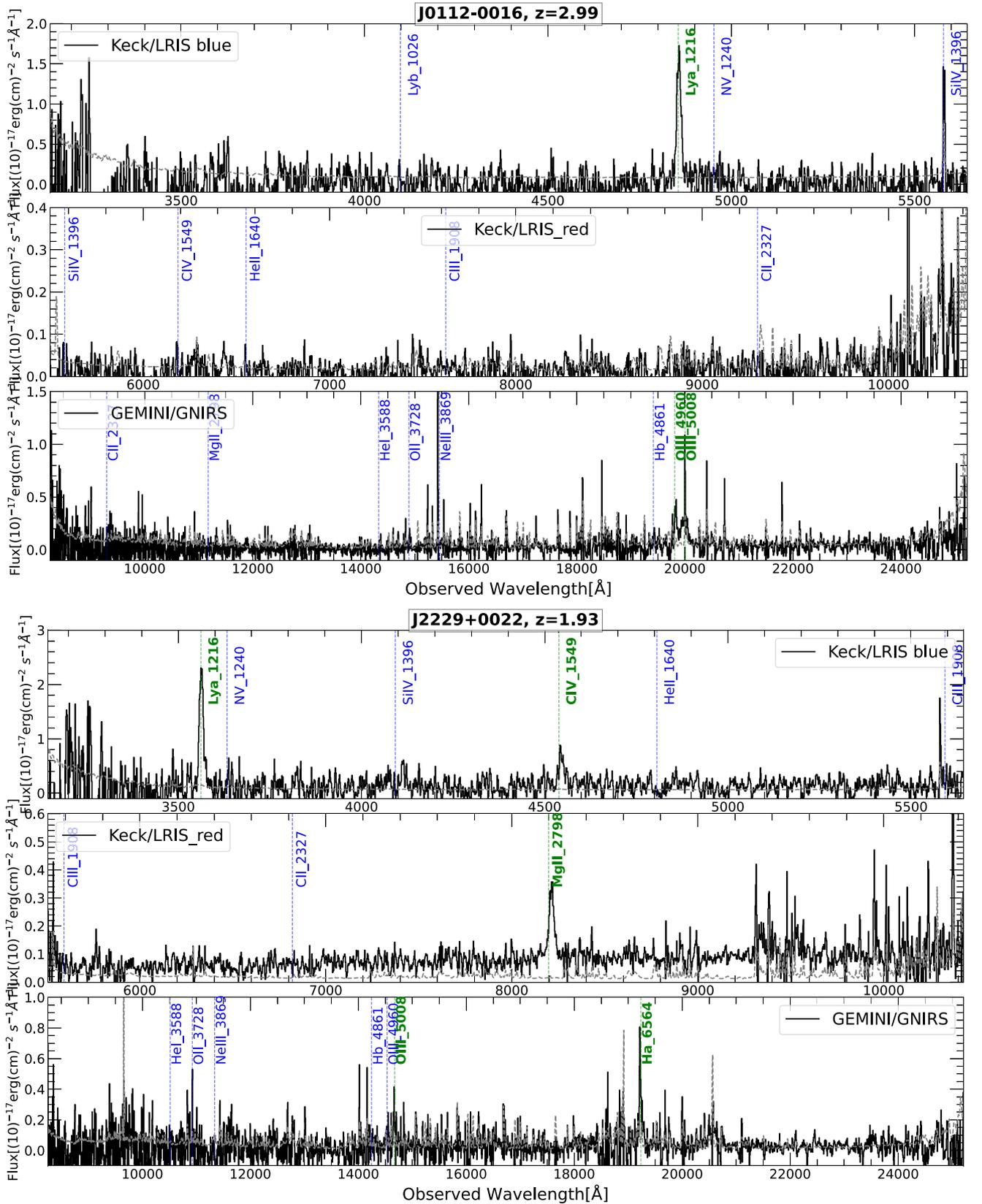

**Figure 3.** Example Keck/LRIS and Gemini/GNIRS spectra for two type 2 QSO targets. The spectra for each target are Keck/LRIS blue, Keck/LRIS red, and Gemini/GNIRS from the top to bottom. The expected emission lines are shown as blue dashed lines, and the detected emission lines are shown in green. The first target J0112−0016 is the highest redshift type 2 QSO in our sample at $z = 2.99$. We detected the Ly $\alpha$ emission line in Keck/LRIS blue and the [O III] emission line in Gemini/GNIRS. The H $\alpha$ emission line is beyond the wavelength coverage of ground-based telescopes and can only be observed from space. The second target J2229+0022 shows the detection of strong narrow UV lines.

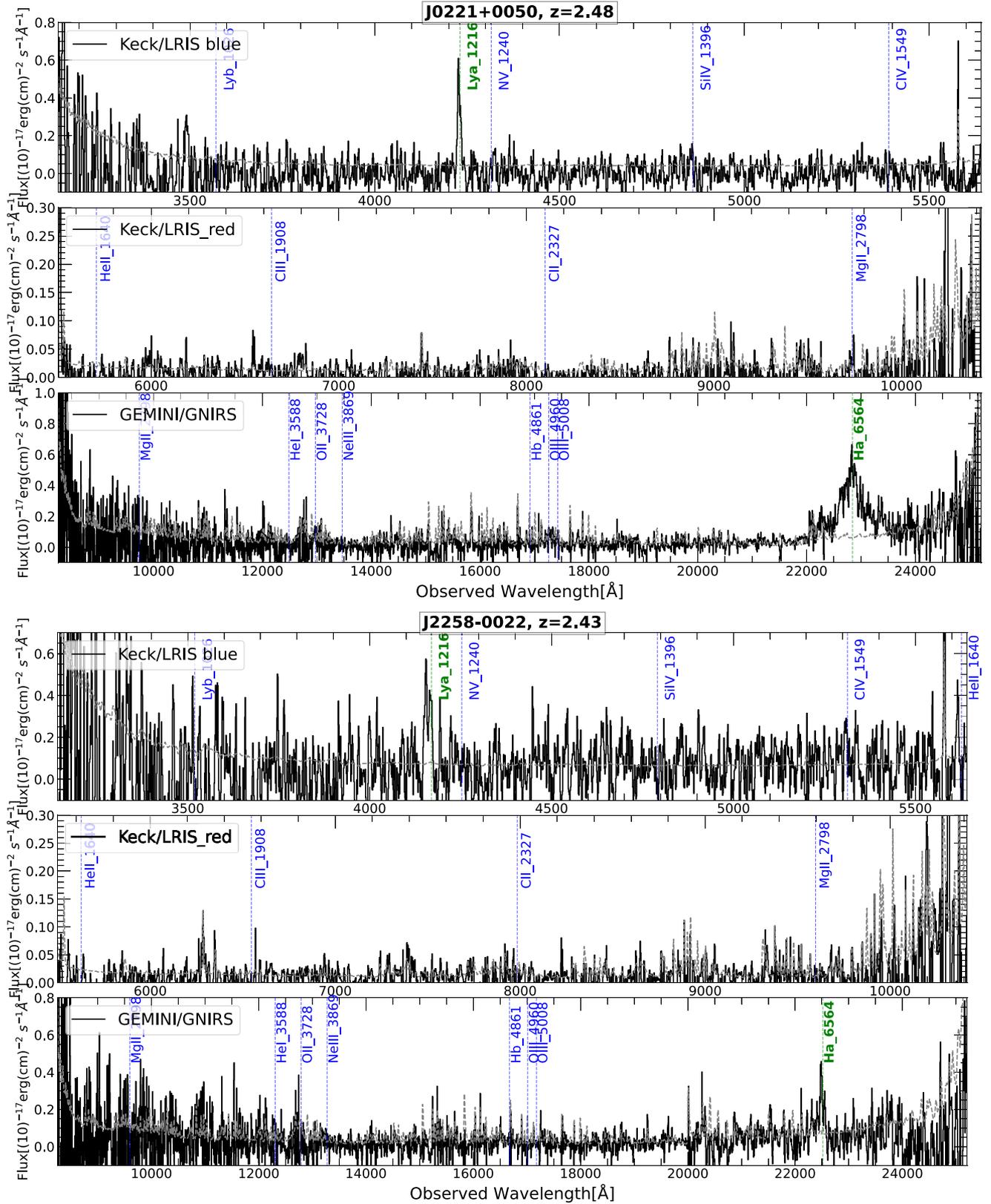

**Figure 4.** Example Keck/LRIS and Gemini/GNIRS spectra for two 'LRD-like' targets. The spectra for each target are Keck/LRIS blue, Keck/LRIS red, and Gemini/GNIRS from the top to bottom. The expected emission lines are shown as blue dashed lines, and the detected emission lines are shown in green. The two targets J0221+0050 and J2258−0022 only have broad H α emission lines and narrow Ly α emission lines. The spectra of our sample show a huge diversity and will be shown in detail in the coming paper.

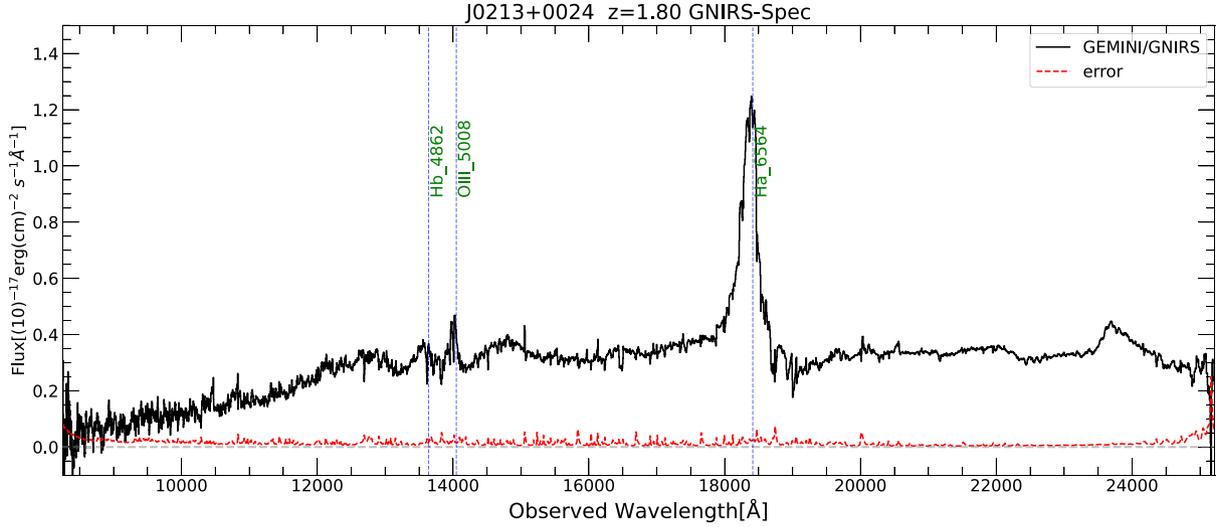

**Figure 5.** GNIRS spectra of the only candidate reddened type 1 QSO in our sample. This target has a strong continuum and broad H $\alpha$ emission line.

**Table 1.** The target name, redshift, and photometry ABmag in each band of all 24 targets.

| Target | Redshift | g | r | i | z | y | Y | J | H | K |
|---|---|---|---|---|---|---|---|---|---|---|
| J0024−0012 | 1.53 | 23.99 ± 0.10 | 23.52 ± 0.06 | 23.81 ± 0.34 | 22.41 ± 0.07 | > 23.34 | > 23.13 | 21.52 ± 0.47 | > 21.42 | > 22.16 |
| J0041−0029 | 2.09 | 24.09 ± 0.11 | 23.76 ± 0.11 | > 25.55 | 22.78 ± 0.13 | > 24.22 | > inf | > 23.52 | > 21.90 | > 21.70 |
| J0047+0003 | – | > 25.43 | 24.07 ± 0.23 | > 25.10 | 23.00 ± 0.26 | > 23.43 | > 24.64 | > 23.18 | > 23.23 | > 22.45 |
| J0054+0047 | 2.17 | 25.49 ± 0.43 | 24.69 ± 0.24 | > 25.05 | 22.67 ± 0.11 | 22.52 ± 0.42 | > 23.05 | > 22.51 | > 22.16 | > 22.01 |
| J0105−0023 | 1.87 | 24.68 ± 0.13 | 24.02 ± 0.09 | 23.88 ± 0.39 | 23.21 ± 0.12 | > 24.10 | > 23.92 | > 23.93 | > 22.93 | > 22.92 |
| J0112−0016 | 2.99 | 25.41 ± 0.30 | 24.60 ± 0.22 | > 24.51 | 24.47 ± 0.37 | > 23.88 | > 24.10 | > 23.12 | > 23.55 | > 23.40 |
| J0113+0029 | 2.33 | 24.55 ± 0.15 | 24.10 ± 0.13 | > 24.99 | 23.46 ± 0.21 | > 22.74 | > 23.11 | > 23.19 | > 22.96 | > 23.11 |
| J0130+0009 | 2.50 | 24.54 ± 0.24 | > 25.56 | > 25.42 | > 24.92 | > 22.78 | > 22.57 | > 23.64 | > 22.50 | > 21.64 |
| J0149+0052 | 1.85 | 24.44 ± 0.09 | 24.00 ± 0.08 | 24.62 ± 0.50 | 23.31 ± 0.13 | > 25.82 | > 22.53 | 21.60 ± 0.51 | 21.22 ± 0.49 | > 22.96 |
| J0150+0056 | 3.49 | 24.02 ± 0.07 | 23.56 ± 0.06 | > 25.44 | 23.56 ± 0.17 | > 23.34 | > 23.83 | > 22.43 | > 23.73 | > 22.57 |
| J0152−0024 | 2.78 | 23.32 ± 0.04 | 22.83 ± 0.04 | 24.05 ± 0.37 | 22.65 ± 0.07 | > 23.69 | > 22.06 | > 22.35 | > 22.02 | 20.98 ± 0.32 |
| J0213+0024 | 1.81 | 23.58 ± 0.06 | 23.13 ± 0.04 | 23.00 ± 0.15 | 21.17 ± 0.02 | 21.56 ± 0.21 | 21.14 ± 0.18 | 19.90 ± 0.08 | 19.33 ± 0.09 | 18.69 ± 0.04 |
| J0214−0000 | 1.63 | 23.19 ± 0.05 | 22.87 ± 0.06 | 23.70 ± 0.27 | 22.09 ± 0.09 | > 23.83 | > 23.27 | 21.73 ± 0.43 | > 22.04 | 21.11 ± 0.40 |
| J0215+0042 | 0.88 | 23.93 ± 0.07 | 22.78 ± 0.03 | 22.77 ± 0.14 | 21.28 ± 0.03 | 21.48 ± 0.22 | 21.38 ± 0.31 | > 23.95 | 20.26 ± 0.21 | 19.22 ± 0.07 |
| J0221+0050 | 2.48 | 24.57 ± 0.21 | 24.44 ± 0.28 | > 24.99 | 23.67 ± 0.40 | > 23.55 | 21.93 ± 0.51 | > 22.46 | > 22.09 | 21.17 ± 0.42 |
| J2229+0022 | 1.93 | 24.18 ± 0.15 | 23.26 ± 0.08 | 23.94 ± 0.32 | 22.37 ± 0.09 | > 22.53 | > 22.27 | 21.64 ± 0.45 | > 21.64 | 20.98 ± 0.38 |
| J2233−0004 | 1.60 | 24.73 ± 0.22 | 23.42 ± 0.08 | 23.56 ± 0.18 | 22.00 ± 0.06 | 21.95 ± 0.40 | > 22.90 | 21.51 ± 0.46 | > 21.44 | 20.62 ± 0.30 |
| J2239−0030 | 1.91 | 24.03 ± 0.08 | 23.43 ± 0.05 | 24.37 ± 0.37 | 22.90 ± 0.09 | > 23.97 | > 23.33 | > 23.70 | > 22.82 | 21.19 ± 0.47 |
| J2239−0054 | 2.09 | 24.02 ± 0.10 | 23.67 ± 0.09 | 24.48 ± 0.45 | 22.51 ± 0.09 | > 22.80 | > 22.72 | > 22.78 | > 22.09 | > 22.31 |
| J2243+0017 | 1.91 | 23.90 ± 0.08 | 23.32 ± 0.05 | 23.58 ± 0.19 | 21.65 ± 0.04 | > 22.59 | > 23.48 | 20.71 ± 0.21 | 19.82 ± 0.16 | 18.92 ± 0.06 |
| J2258−0022 | 2.42 | 24.31 ± 0.11 | 23.96 ± 0.10 | > 26.42 | 23.16 ± 0.13 | > 24.28 | > 23.81 | > 22.97 | > 22.53 | > 22.61 |
| J2259−0009 | 1.89 | 23.58 ± 0.07 | 23.22 ± 0.07 | 23.47 ± 0.24 | 22.26 ± 0.08 | > 24.30 | > 22.54 | 22.25 ± 0.49 | 21.38 ± 0.38 | 20.39 ± 0.18 |
| J2329+0020 | 2.67 | 24.25 ± 0.13 | 23.91 ± 0.11 | > 24.35 | 22.92 ± 0.12 | > 22.96 | > 24.87 | > 22.84 | 21.17 ± 0.43 | > 23.38 |
| J2334+0031 | 2.10 | 23.00 ± 0.03 | 22.92 ± 0.04 | 23.09 ± 0.16 | 22.34 ± 0.05 | > 24.50 | 22.07 ± 0.45 | > 23.72 | > 22.52 | > 22.05 |

*Cold dust.* The model is S17 from Schreiber et al. (2018) with very small grains (VSG) correction. This model consists of two independent components: the dust continuum and the mid-IR emission line spectra from complex molecules. The free parameters in this model are: T_dust (the cold dust temperature 14–42 K), fracPAH [polycyclic aromatic hydrocarbons (PAHs) fraction], and the normalization parameter SB.

We take the (Ly $\alpha$ forest corrected) photometry from PS1, DESI Legacy Survey, UKIDSS, *Spitzer*, *UNWISE*, and *ALLWISE* for all 23 identified targets as the input of the AGNFITTER. The pipeline runs a 10 000-step Markov chain Monte Carlo (MCMC) to fit the four templates to the measured photometry. The pipeline gives the fitted template, free parameters listed above, and other derived physical properties [like stellar mass, star formation rate (SFR) in optical and IR, luminosity, and AGN fraction in different wavelength ranges] as the outputs. The formula to calculate the SFR in this pipeline is presented in Section 4.4.

In the SED fitting parameter set, we choose the PRIOR_AGNfraction and turn_on_AGN to be true. Setting PRIOR_AGNfraction to be true is to give preference to AGN contribution in the UV and optical if the blue/UV bands are 10 times higher than expected by the galaxy luminosity function in Parsa et al. (2016). Setting turn_on_AGN to be true is to include the accretion disc and dust torus components in the fitting. We set `PRIOR_energy_balance` to be `Flexible`, which means that the model favours a combination of parameters such that the luminosity of the cold dust and that attenuated in the stellar component are equal.

**Table 2.** The target name, redshift, photometry ABmag in each band, and the RA, Dec. of all 24 targets.

| Target | Redshift | W1 | IRAC1 | IRAC2 | W2 | W3 | W4 | RA | Dec. |
|---|---|---|---|---|---|---|---|---|---|
| J0024−0012 | 1.53 | 18.97 ± 0.07 | 18.44 ± 0.01 | 17.43 ± 0.01 | 17.43 ± 0.04 | 15.65 ± 0.08 | 14.55 ± 0.21 | 00:24:07.02 | −00:12:37.2 |
| J0041−0029 | 2.09 | 19.89 ± 0.15 | 20.10 ± 0.06 | 19.33 ± 0.03 | 18.79 ± 0.11 | 16.30 ± 0.13 | 14.28 ± 0.165 | 00:41:57.77 | −00:29:32.1 |
| J0047+0003 | – | 20.12 ± 0.19 | 19.05 ± 0.23 | 18.98 ± 0.19 | 19.05 ± 0.17 | 15.23 ± 0.07 | 13.81 ± 0.13 | 00:47:29.20 | +00:03:59.2 |
| J0054+0047 | 2.17 | 19.63 ± 0.11 | 19.46 ± 0.03 | 18.57 ± 0.02 | 18.80 ± 0.15 | 15.92 ± 0.13 | 14.39 ± 0.16 | 00:54:24.45 | +00:47:50.2 |
| J0105−0023 | 1.87 | 20.32 ± 0.20 | 19.94 ± 0.11 | 18.59 ± 0.04 | 18.46 ± 0.09 | 15.19 ± 0.06 | 14.33 ± 0.20 | 01:05:52.86 | −00:23:51.2 |
| J0112−0016 | 2.99 | 20.70 ± 0.30 | 20.32 ± 0.39 | 19.76 ± 0.16 | 20.23 ± 0.39 | 16.60 ± 0.19 | 14.45 ± 0.19 | 01:12:22.64 | −00:16:33.0 |
| J0113+0029 | 2.33 | 21.00 ± 0.40 | 20.32 ± 0.20 | 19.15 ± 0.06 | 19.63 ± 0.24 | 15.90 ± 0.09 | 14.58 ± 0.20 | 01:13:14.49 | +00:29:17.1 |
| J0130+0009 | 2.50 | 19.63 ± 0.10 | 19.16 ± 0.06 | 18.28 ± 0.03 | 18.28 ± 0.07 | 15.68 ± 0.07 | 14.55 ± 0.17 | 01:30:33.47 | +00:09:50.4 |
| J0149+0052 | 1.85 | 18.97 ± 0.06 | 18.61 ± 0.02 | 17.70 ± 0.01 | 17.61 ± 0.05 | 15.57 ± 0.07 | 14.58 ± 0.21 | 01:49:39.96 | +00:52:56.7 |
| J0150+0056 | 3.49 | 19.01 ± 0.06 | 20.02 ± 0.06 | 19.96 ± 0.06 | 19.02 ± 0.12 | 16.05 ± 0.09 | 13.86 ± 0.09 | 01:50:55.28 | +00:56:00.2 |
| J0152−0024 | 2.78 | 18.65 ± 0.05 | 18.75 ± 0.02 | 18.06 ± 0.01 | 17.92 ± 0.05 | 15.47 ± 0.06 | 14.46 ± 0.16 | 01:52:35.29 | −00:24:59.4 |
| J0213+0024 | 1.81 | 17.53 ± 0.03 | 17.25 ± 0.01 | 16.66 ± 0.01 | 16.70 ± 0.03 | 15.15 ± 0.05 | 14.16 ± 0.12 | 02:13:45.44 | +00:24:36.1 |
| J0214−0000 | 1.63 | 19.55 ± 0.09 | 19.94 ± 0.09 | 18.63 ± 0.02 | 18.44 ± 0.08 | 15.48 ± 0.06 | 14.42 ± 0.14 | 02:14:26.98 | −00:00:21.3 |
| J0215+0042 | 0.88 | 17.67 ± 0.03 | 17.34 ± 0.01 | 16.69 ± 0.01 | 16.71 ± 0.03 | 15.07 ± 0.05 | 13.99 ± 0.12 | 02:15:14.76 | +00:42:23.8 |
| J0221+0050 | 2.48 | 18.82 ± 0.06 | 18.88 ± 0.02 | 18.21 ± 0.01 | 17.99 ± 0.05 | 15.82 ± 0.08 | 14.30 ± 0.14 | 02:21:27.60 | +00:50:24.6 |
| J2229+0022 | 1.93 | 18.64 ± 0.06 | 19.49 ± 0.04 | 19.16 ± 0.03 | 18.35 ± 0.08 | 15.11 ± 0.06 | 13.64 ± 0.11 | 22:29:20.83 | +00:22:53.5 |
| J2233−0004 | 1.60 | 19.60 ± 0.12 | 19.42 ± 0.04 | 18.98 ± 0.02 | 18.89 ± 0.12 | 16.55 ± 0.18 | 14.42 ± 0.19 | 22:33:58.38 | −00:04:14.9 |
| J2239−0030 | 1.97 | 20.15 ± 0.18 | 20.31 ± 0.07 | 19.10 ± 0.02 | 19.17 ± 0.16 | 15.25 ± 0.06 | 14.07 ± 0.15 | 22:39:04.01 | −00:30:54.9 |
| J2239−0054 | 2.09 | 19.41 ± 0.10 | 19.64 ± 0.04 | 18.62 ± 0.03 | 18.46 ± 0.09 | 15.60 ± 0.08 | 14.35 ± 0.18 | 22:39:11.98 | −00:54:22.3 |
| J2243+0017 | 1.91 | 17.66 ± 0.04 | 17.51 ± 0.01 | 17.06 ± 0.01 | 17.12 ± 0.04 | 15.35 ± 0.08 | 14.11 ± 0.17 | 22:43:38.04 | +00:17:49.9 |
| J2258−0022 | 2.42 | 21.27 ± 0.50 | 20.08 ± 0.05 | 19.14 ± 0.03 | 19.16 ± 0.16 | 16.47 ± 0.16 | 14.49 ± 0.20 | 22:58:51.90 | −00:22:07.0 |
| J2259−0009 | 1.89 | 18.74 ± 0.06 | 18.38 ± 0.01 | 17.55 ± 0.01 | 17.55 ± 0.05 | 15.54 ± 0.09 | 14.36 ± 0.18 | 22:59:56.84 | −00:09:18.4 |
| J2329+0020 | 2.67 | >21.47 | 20.37 ± 0.08 | 19.95 ± 0.04 | 20.33 ± 0.43 | 16.34 ± 0.15 | 14.48 ± 0.19 | 23:29:25.01 | +00:20:57.7 |
| J2334+0031 | 2.10 | 19.36 ± 0.09 | 19.40 ± 0.03 | 18.55 ± 0.02 | 18.46 ± 0.09 | 15.94 ± 0.10 | 14.23 ± 0.16 | 23:34:41.49 | +00:31:14.0 |

**Table 3.** The band name, central wavelength, and bandwidth used in this paper.

| Survey | Band | Central wave (μm) | Bandwidth (μm) |
|---|---|---|---|
| DESI Legacy | g | 0.476 | 0.238 |
| DESI Legacy | r | 0.668 | 0.212 |
| DESI Legacy | z | 0.863 | 0.239 |
| Pan-STARRS1 Survey (PS1) | i | 0.755 | 0.152 |
| Pan-STARRS1 Survey (PS1) | y | 0.963 | 0.178 |
| UKIRT Infrared Deep Sky Survey (UKIDSS) | Y | 1.03 | 0.102 |
| UKIRT Infrared Deep Sky Survey (UKIDSS) | J | 1.25 | 0.159 |
| UKIRT Infrared Deep Sky Survey (UKIDSS) | H | 1.64 | 0.292 |
| UKIRT Infrared Deep Sky Survey (UKIDSS) | K | 2.12 | 0.351 |
| *Spitzer* | IRAC1 | 3.6 | 0.832 |
| *Spitzer* | IRAC2 | 4.5 | 1.14 |
| *WISE* | W1 | 3.3 | 0.66 |
| *WISE* | W2 | 4.6 | 1.04 |
| *WISE* | W3 | 11.5 | 5.51 |
| *WISE* | W4 | 22 | 4.10 |

### 3.4 SED fitting results

Two examples showing cut-outs and SED fitting results are shown in Fig. 6. Results for the rest of the targets are shown in Appendix B. The upper panel of each figure shows the image cut-outs with a size of 60 arcsec and a 3 arcsec red circle is placed at the centre of each cut-out, indicating the location of the source. The bottom panel of the figures shows the photometry and fitted SED templates. The photometry from different surveys is coloured in groups: DESI Legacy Survey (blue), PS1 (cyan), UKIDSS (magenta), *Spitzer* (purple), and *WISE* (orange). The maximum likelihood total fitted SED model is shown by the black solid line. The different components are also plotted: the accretion disc in dark blue, the stellar emission in green, and the dust torus in red. The solid line is the maximum likelihood of each component, and 100 models constructed from combinations of parameters randomly selected from the posterior are plotted as a shaded area.

Since we set `PRIOR_AGNfraction` and `turn_on_AGN` both to be true, the pipeline gives the AGN component (accretion disc and dust torus) a high priority (relative to the stellar emission and cold dust) to fit. The hot dust torus templates are fitted well to the *Spitzer* and *WISE* photometry in most cases, so the starburst component with cold dust does not contribute much to the fitting results. J0105−0023 and J0130+0009 show strong cold dust contribution in the fitting results. The J0130+0009 target has the highest cold dust contribution among the 23 targets, the luminosity contribution from the cold dust at around 100 μm is about 0.9. The cold dust contributes 0.39 luminosity in J0105−0023, at around 100 μm. The cold dust contributions in the rest of the 21 targets are all below 0.01, so the dust torus alone can explain the mid-IR emission. We test how the fitting results change without the AGN hot dust torus contribution in Section 4.4. The AGN fractions for all targets are all above 0.99 in 8–1000 μm, indicating these targets all have not hot dust torus components and significant AGN contributions.

One target J0113+0029 has a very 'clean' AGN contribution: only the accretion disc and hot dust torus components show up in the SED fitting, and the AGN fractions at all wavelength bins are above 0.99. The remaining 22 targets all have stellar emission components that contribute to the fitting results. The rest-UV/optical fits are a degenerate between the QSO and galaxy; the fits here give just one possible scenario: the accretion disc contributes to the rest-UV and the stellar emission from the galaxy dominates at around 1 µm. However, for five targets (J0152−0024, J0221+0050, J0213+0024, J2243+0017, and J2259−0009), the pipeline fits an extremely bright stellar emission component (above $10^{46}$ erg s$^{-1}$, demanding stellar mass above $10^{12}$ M$_\odot$), which is an unphysical solution. As such, we assign an upper limit to the stellar mass and rerun the fitting for these five targets. In this case, the pipeline fits a reddened accretion disc component around 1 µm and a stellar emission galaxy component to the rest-UV. For the remaining 17 targets, AGNFITTER applies a reddened accretion disc template to fit the rest-UV. However, this pipeline does not enforce the energy balance between the accretion disc and the hot dust. While we consider the hot dust torus fitting to be robust, the reddened accretion disc component remains uncertain due to the lack of energy balance.

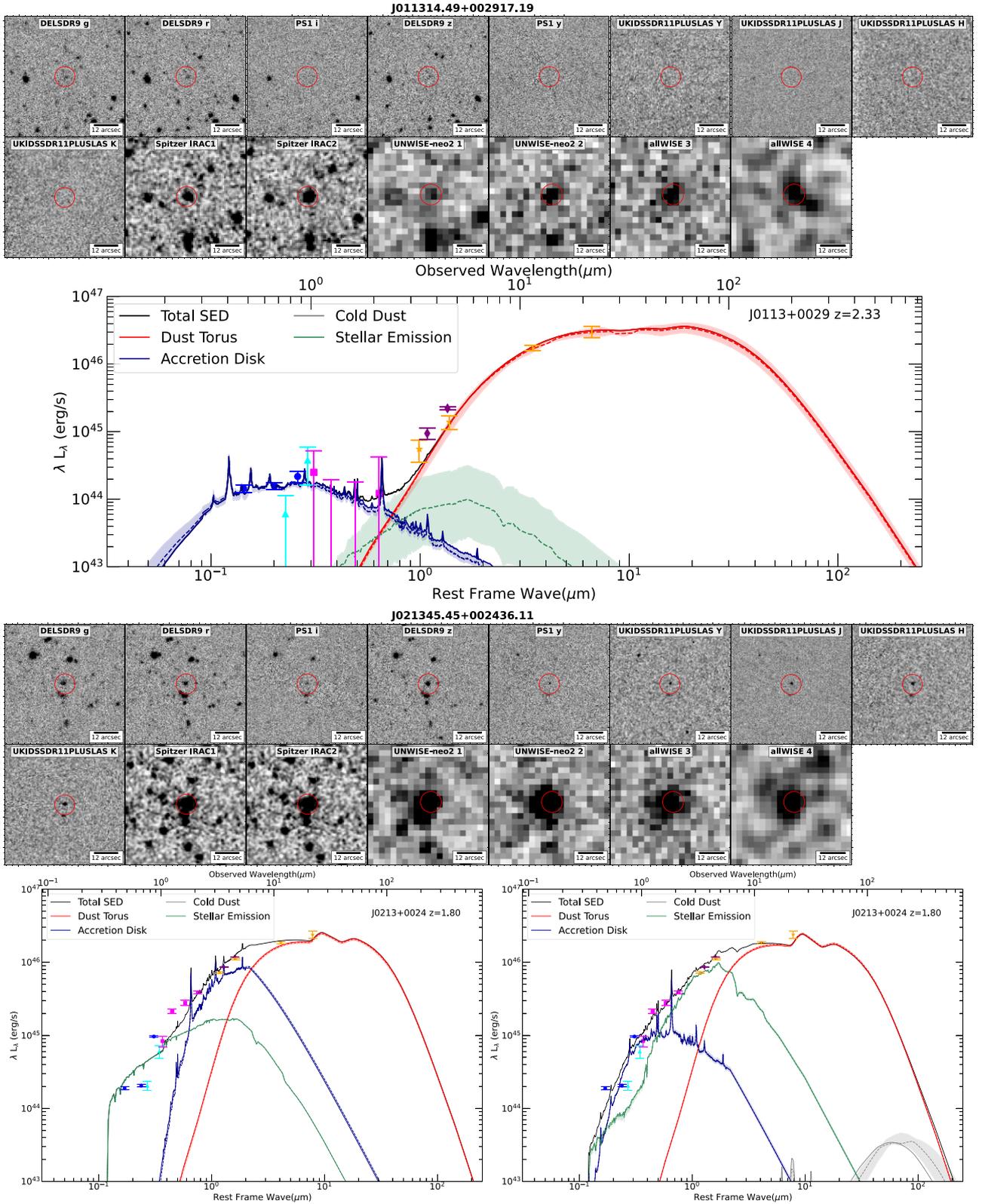

**Figure 6.** Example cut-outs and SED fitting result for two targets. Each upper panel shows the cut-out from DESI Legacy Survey (*g*, *r*, *z*), PS1 (*i*, *y*), UKIDSS (*Y*, *J*, *H*, *K*), *Spitzer* (IRAC1, IRAC2), and *WISE* (*W*1, *W*2, *W*3, *W*4). Each lower panel shows the SED fitting result from the AGNFITTER. The photometry data from DESI Legacy Survey, PS1, UKIDSS, *Spitzer*, and *WISE* are shown as blue, cyan, pink, purple, and orange data points. The coloured solid lines are max likelihood fit of the four components from different templates. The coloured dashed line is the median fitting from 100 models constructed randomly using the posterior and the shadow region is the 25th–75th percentile. The black solid line is the total SED fitting result. The J0113+0029 SED is fitted by the dust torus and the accretion disc. We give two fittings for target J0213+0024 (the second one seems unphysical considering the extreme brightness of the galaxy component). The IR emission is dominated by the dust torus, and the rest-UV/optical is a mixture of QSO and galaxy light.

Our main purpose in this paper is to investigate the components in the composite SED to better understand this population, instead of discussing the individual targets in detail. The rest-frame UV/optical light is a degenerate between the accretion disc and the stellar emission components; we will discuss this in detail in Section 4. The fitted results are mainly used for the photometry normalization and reveal the AGN contribution.

### 3.5 Bolometric luminosity

The SED fitting reveals the hot dust torus component in each target. Here, we define the luminosity of the torus to be

$$L_{torus} = \int_{1\,\mu m}^{1000\,\mu m} L_\lambda \, d\lambda. \tag{3}$$

According to the SED results, the torus luminosity of our targets is $L_{torus} = 10^{46.53}$–$10^{47.15}$ erg s$^{-1}$. The bolometric luminosity can be defined as in Lusso et al. (2013):

$$R \equiv \frac{L_{torus}}{L_{bol}} = \frac{\int_{1\,\mu m}^{1000\,\mu m} L_\lambda \, d\lambda}{\int_{\lambda_{min}}^{1\,\mu m} L_\lambda^* \, d\lambda}, \tag{4}$$

where $R$ is the obscured fraction. The bolometric luminosity defined here represents the optical–UV and X-ray emission emitted by the nucleus and reprocessed by the dust grains in the torus. The $L_{torus}$ can be integrated using the fitted torus template. The $L_\lambda^*$ is the intrinsic luminosity. Due to the dust obscuration, the $L_{bol}$ cannot be measured directly but can be estimated by assuming a value of $R$. There are two extreme values for the estimate of obscured fraction: (1) the obscured fraction is 0.5 which means the number density of type 1 and type 2 QSOs is comparable; (2) the obscured fraction is 1, which means all the QSOs are obscured. According to our number density estimate (see Section 2) and some other searches (Polletta et al. 2007; Merloni et al. 2014; Lusso et al. 2015), we estimate the obscured fraction to be 0.5. Under this assumption, the bolometric luminosities of our targets have a range of $L_{bol} = 10^{46.28}$–$10^{48.08}$ erg s$^{-1}$. We have listed the bolometric luminosity for all the targets in Table B2. The median value is $L_{bol} = 10^{47.04}$ erg s$^{-1}$. Assuming an Eddington limit, these targets have black hole masses in the range of $10^{8.18}$–$10^{9.98}$ M$_\odot$. Ishikawa et al. (2023) also report the bolometric luminosity of these targets using an IR-bolometric correction:

$$L_{torus} = 8 \times L_{3.45\,\mu m}. \tag{5}$$

The bolometric luminosity using this equation is $L_{bol} = 10^{46.23}$–$10^{47.72}$ erg s$^{-1}$ (shown in Table B2).

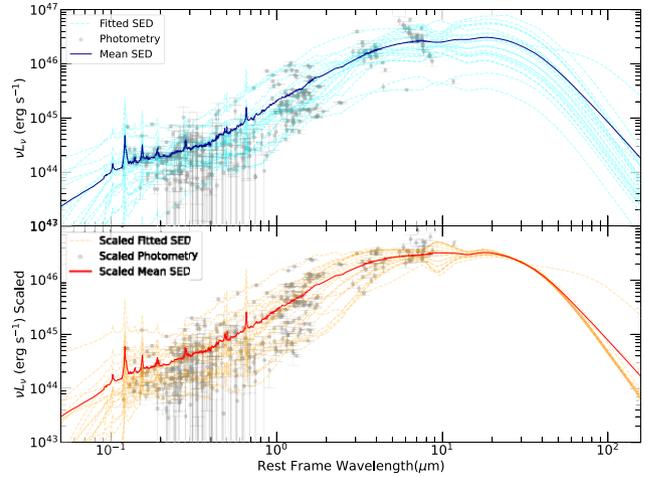

**Figure 7.** The scaled SED fitting results for all targets in our sample. Every target is scaled to $\lambda L_\lambda = 10^{46.33}$ erg s$^{-1}$ (the mean $\lambda L_\lambda$ at 30 µm value in our sample). Then the same scaling parameter is applied to the photometry. The top panel shows the unscaled SED fitting for individual targets in dashed cyan. The mean SED is shown in solid dark blue. The scaled photometry, fitted SED, and mean SED are shown in grey, orange, and red in the lower panel.

### 3.6 Composite photometry

After obtaining SED model fits for each target, we use this fitted model to normalize the photometry of all the targets. We use the luminosity at the rest-frame 30 µm to scale the SED model of all the targets. The rest-frame 30 µm is chosen because there are no emission or absorption lines at that wavelength. Every target is scaled to $\lambda L_\lambda = 10^{46.33}$ erg s$^{-1}$ (the mean $\lambda L_\lambda$ at 30 µm value in our sample). Then we take the same scaling parameter to scale the photometry for all the targets. The result is shown in Fig. 7. The upper panel shows the unscaled SED, fitted SED for all targets, and the mean SED. The lower panel shows the scaled photometry, fitted SED for all targets, and the mean SED.[2]

The scaled photometry for all the targets and the composite photometry are shown in Fig. 8. The photometric measurements from different imaging surveys are grouped in colours: PS1 in blue, DESI Legacy Survey in cyan, UKIDSS in pink, *Spitzer* in purple, and *WISE* in orange. The different markers indicate different targets.

The photometric measurements are divided into 20 wavelength bins on a logarithmic scale. The mean values of luminosity are taken in each bin to make the composite photometry. The final composite photometry is plotted as red stars in Fig. 8.

---

[2] We notice the targets show different features (emission or absorption) at the rest-frame 10 µm. This feature is related to the silicate emission in QSOs. Siebenmorgen et al. (2005) claim that due to the dust torus around the central AGN, the 10 µm silicate should be seen as absorption in type 2 QSOs but as emission lines in type1 QSOs. In our SED fitting results, the reddened type 1 QSOs show emission features at 10 µm, while 11 of 22 type 2 QSOs show absorption features at 10 µm. However, we do not consider this 10 µm feature as an observational result since: (1) we do not have real measurements at 10 µm, thus the fitting result is strongly model dependent; (2) the reddened type 1 QSOs still have faint optical magnitudes and very red colours, making them slightly different from typical type 1 QSOs. So, how to group these targets will still affect the analysis. But we still take this as an interesting result because detecting such a silicate feature will help us better understand the structure of the dust torus around AGN.

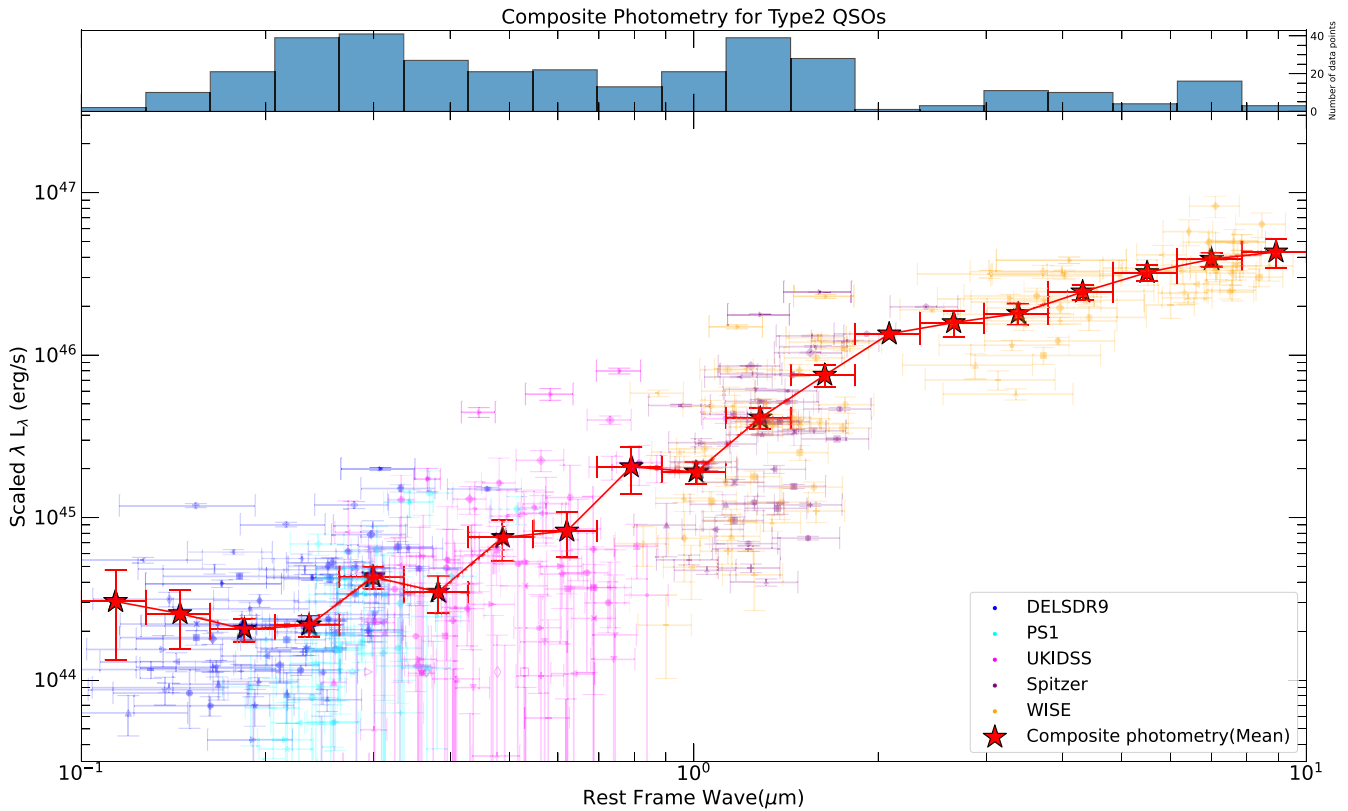

**Figure 8.** The scaled composite photometry results for all type 2 QSOs. All the measured photometric data points are scaled using the same scaling parameter of the fitted templates. The scaled data are shown in different colours: DESI in blue, PS1 in cyan, UKIDSS in magenta, *Spitzer* in purple, and *WISE* in orange. 20 bins are shown at a log scale, and the upper subplot indicates the number of data points in each bin. The mean photometric values of each bin are shown as red stars.

The upper panel in Fig. 8 shows the number of data points in each bin. Some bins do not have many data points due to the redshift and the wavelength coverage of the survey bands. But most bins have more than 10 measurements. The error bars shown are the errors on the mean value in each bin, specifically if this is computed by

$$\sigma_{\bar{x}} = \frac{\sigma_x}{\sqrt{N}} = \sqrt{\frac{1}{N(N-1)} \sum_{i=1}^{N} (x_i - \bar{x})^2}. \quad (6)$$

### 3.7 Comparison to previous type 2 quasar SEDs

Here we compare our composite SED to the SED from type 2 AGN in SDSS (Hickox et al. 2017), *Spitzer* Wide-Area Infrared Extragalactic survey (SWIRE; Polletta et al. 2007), and the SED of Hot DOGs (Fan et al. 2016). The *Hickox/Reyes* SED is generated using the type 2 QSO sample in Reyes et al. (2008). These type 2 QSOs are selected from the SDSS spectroscopic data base with narrow emission lines full width at half-maximum (FWHM) < 1000 km s⁻¹. The type 2 QSOs from the SED in *Polletta/SWIRE* are obtained by combining the observed optical/near-IR spectrum of the QSO sample in Gregg et al. (2002) and Polletta et al. (2006). The *Hot DOGs* SED are generated using the targets selected from a 'W1W2 drop' selection (Fan et al. 2016).

We take these arbitrarily normalized published SEDs and scale them to the typical bolometric luminosity of sources from they were determined. Specifically, the bolometric luminosity of the sources studied by Hickox et al. (2017, *Hickox/Reyes*) was estimated using the luminosity of the [O III] lines, $L_{\text{O III}}$. These type 2 QSOs have a median $L_{\text{O III}} \sim 10^9 \, L_\odot = 10^{42.58} \, \text{erg s}^{-1}$. Adopting the correction $L_{\text{bol}} \sim 912 \times L_{\text{O III}}$ from Lamastra et al. (2009), the bolometric luminosity is $L_{\text{bol}} \sim 10^{45.54} \, \text{erg s}^{-1}$. The bolometric luminosity of type 2 AGNs in Polletta et al. (2007) is about $L_{\text{bol}} \sim 10^{45.5} \, \text{erg s}^{-1}$. The median SED of Hot DOGs in Fan et al. (2016) has $L_{\text{IR}} \sim 10^{46.58} \, \text{erg s}^{-1}$, and the bolometric luminosity is computed by $L_{\text{bol}} = 1.4 \times L_{\text{IR}} = 10^{46.73} \, \text{erg s}^{-1}$. Our composite SED has bolometric luminosity $L_{\text{bol}} = 10^{47.04} \, \text{erg s}^{-1}$ (we scaled $\lambda L_\lambda$ at 30 μm to the average value of our sources). The comparison result is shown in Fig. 9.

The SED shapes from Hickox et al. (2017) and Polletta et al. (2007) are similar, but are different from our SED. Because their type 2 QSOs are low-z sources (Hickox et al. 2017) or selected from small area surveys (Polletta et al. 2007), and are hence much fainter intrinsically. Since Galaxy and AGN contribute comparably in fainter sources, their SEDs do not appear red. Our type 2 SED has a much brighter dust torus component (~21×) than these two samples. In our very bright sources, the QSO completely outshines the galaxy, making our SED far redder. The Hot DOG SED has a similar hot dust torus component as in our sample and comparable bolometric luminosity. But these targets are rarer red sources and are selected to be very red (with W1/W2 dropout).

We also show the type 1 QSO SED from Krawczyk et al. (2013). This SED is generated using 119 652 luminous broad-lined QSOs at 0.064 < z < 5.46. The blue dashed line is the mean SED for a subsample of high-luminosity QSOs with $\log(\nu L_\nu)_{\lambda=2500\,\text{Å}} > 45.85$. For the type 1 QSOs, the BBB dominates the rest-UV/optical light, showing an obvious difference to our type 2 QSOs. The SEDs for the ERQs from Hamann et al. (2017) are shown in red. The ERQs have similar bolometric luminosity to our targets, but our targets show a redder colour compared to these ERQs in SDSS.

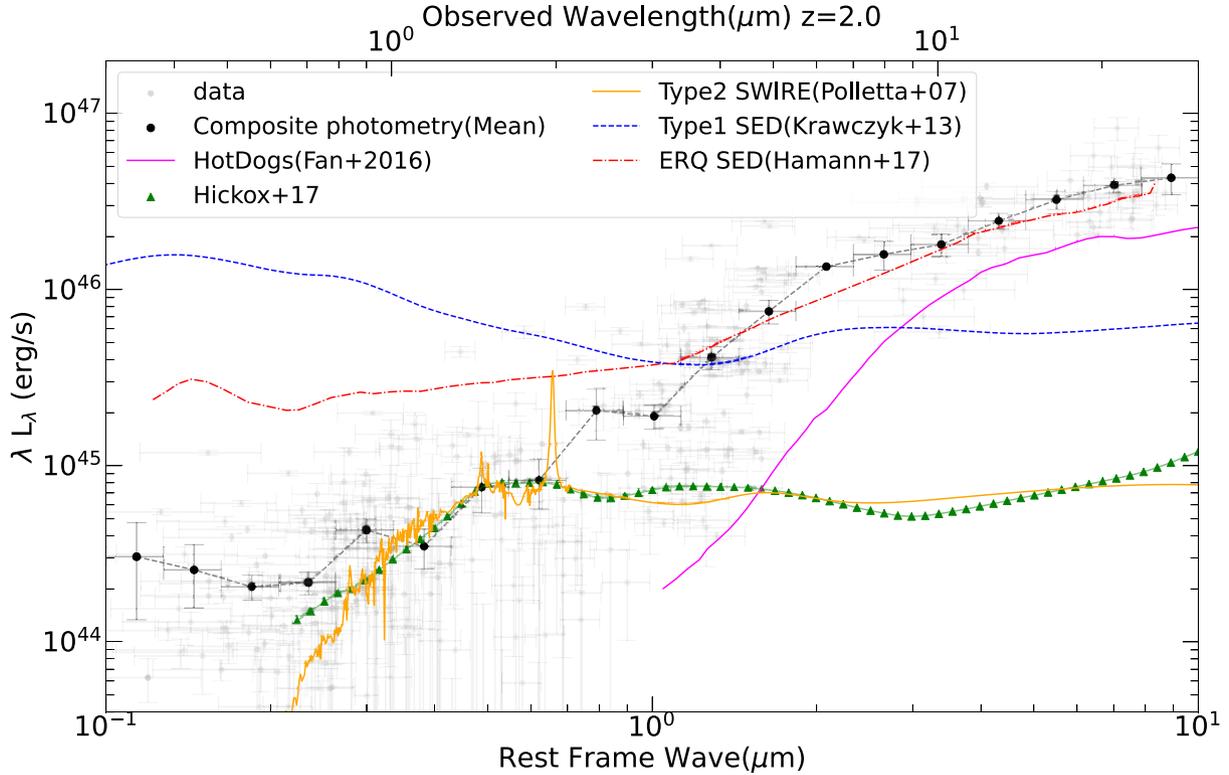

**Figure 9.** SEDs from different surveys. The grey and black points are the normalized photometry and composite photometry in this paper. The green lines show the type 2 SED from the SDSS sample (Hickox et al. 2017), and the SWIRE type 2 SED from Polletta et al. (2007) is shown in orange. Our composite SED shows more reddening and 21 times brighter dust torus than these two samples. In our very bright sources, the QSO completely outshines the galaxy, making our SED far redder. The SED from the Hot DOG sample in Fan et al. (2016) is shown in magenta, and it has a similar hot dust torus component to our targets. But these targets are rarer red sources and are selected to be very red (with $W1/W2$ dropout). We also plot the mean type 1 QSO SED from Krawczyk et al. (2013) in blue and the SED of ERQs from Hamann et al. (2017) in red. For the type 1 QSOs, the BBB dominates the rest-UV/optical light, showing an obvious difference to our type 2 QSOs. Our targets have a redder SED shape even than the ERQs from SDSS.

## 4 DISCUSSION

In this section, we consider the physical origin of our composite type 2 QSO SED with both galaxy and AGN components. We reveal the physical properties that contribute to the luminosity at different wavelengths. For the IR emission, we prefer our targets are dominated by the dust torus. For the rest-UV/optical light, we concluded that the galaxy and AGN contribution are equally possible. Finally, we compare our SED results to *JWST* LRDs. We find similarities on the rest-frame UV SED and discuss the possibility that LRDs can have hot dust torus.

### 4.1 Dust torus and stellar emission in our SED

AGNFITTER yields a fit for the dust torus, accretion disc, and stellar emission component for every target. We take the dust torus as the robust fitting and use it to normalize the photometry to make the composite SED. We plot the mean of the dust torus fitting in

Fig. 10. The grey data points are the scaled photometry and the black dots are our mean composite. The photometric data points are scaled using the same scaling parameter as scaling the dust torus fits (see Section 3.6). The dust torus contributes to the observed IR emission with a mean luminosity $L_{\text{torus}} \sim 10^{46.84}\,\text{erg s}^{-1}$.

We take the galaxy template from Bruzual & Charlot (2003) and quantitatively fit our composite SED around rest-frame 1 μm. If the rest-frame 1 μm is dominated by stellar emission from the galaxy, then the galaxy should have mean mass $M_* \sim 10^{11}\,\text{M}_\odot$. Our targets have black hole mass $10^{8.18}$–$10^{9.98}\,\text{M}_\odot$ assuming the Eddington limit. According to the black hole mass and the host galaxy stellar mass relation (Reines & Volonteri 2015; Suh et al. 2020), a $M_* \sim 10^{11}\,\text{M}_\odot$ host galaxy is reasonable. Moreover, galaxies with $M_* \sim 10^{11}\,\text{M}_\odot$ at $z \sim 2$ have number densities of about $10^{-4}\,\text{dex}^{-1}\,\text{Mpc}^{-3}$ (McLeod et al. 2021; Weaver et al. 2023), which is higher than the number density of QSOs (with $L_{\text{bol}} \sim 10^{47}\,\text{erg s}^{-1}$): $10^{-6}\,\text{dex}^{-1}\,\text{Mpc}^{-3}$ (Shen et al. 2020), indicating that these luminous QSOs can be hosted by a massive galaxy.

For the rest-frame UV light, AGNFITTER yields a good fit when using the accretion disc template with mean $E(B - V) \sim 0.33$. However, the AGNFITTER pipeline only gives an energy balance between the stellar emission in the optical/UV and the reprocessed emission by cold dust in the IR. It does not assume any energy balance between the accretion disc and the dust torus. In the next section, we consider whether the rest-frame UV–optical light can be explained by the reddening of the BBB by invoking energy balance.

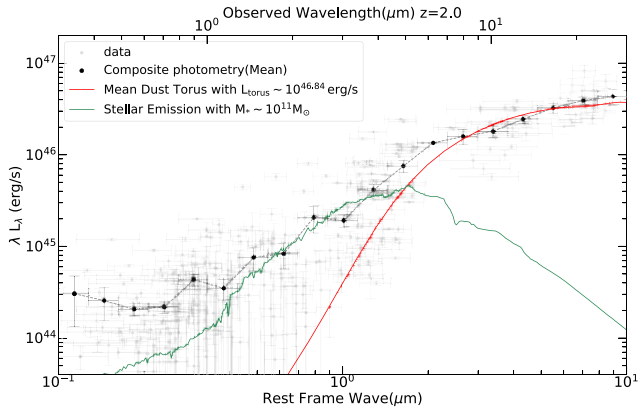

**Figure 10.** The dust torus and stellar emission in our composite SED. The grey data points are the scaled photometries, and the black dots are our mean composite photometry. The dust torus component is the mean value of all the individual fitting results. The dust torus contributes to the observed IR emission with a mean luminosity $L_{torus} \sim 10^{46.84} \, erg \, s^{-1}$. If the rest-frame 1 μm is dominated by the stellar emission, then the galaxy should have mean mass $M_* \sim 10^{11} \, M_\odot$.

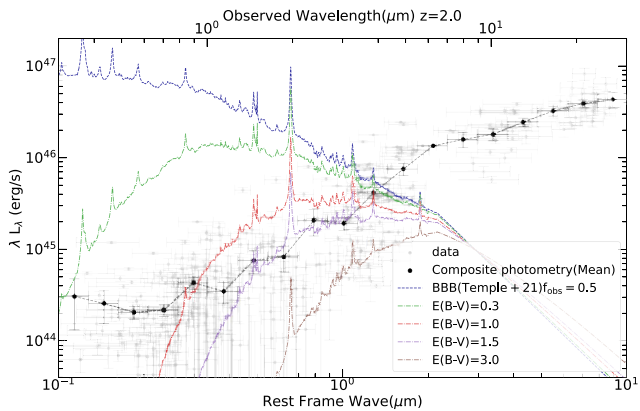

**Figure 11.** The fitting of using different $E(B - V)$ on the BBB template from Temple et al. (2021). The black data points are our mean composite photometry. The blue dashed line is the scaled BBB template assuming an obscured fraction equals 0.5. The coloured dot–dashed lines are the reddened BBB models with different $E(B - V)$ values. None of the reddening models fit our rest-UV data. Some models with $E(B - V) \sim 1.5$ fit our rest-optical data. The reddening BBB light may contribute to our rest-optical data, but cannot explain our rest-UV light.

### 4.2 Could the rest-frame UV–optical light arise from a reddened big blue bump?

Under the assumption of energy balance between the accretion disc and the dust torus, we need to determine how bright the unobscured BBB is. The obscured fraction can be defined as (see Lusso et al. 2013)

$$R = f_{obscuration} = \frac{L_{torus}}{L_{BBB}}. \qquad (7)$$

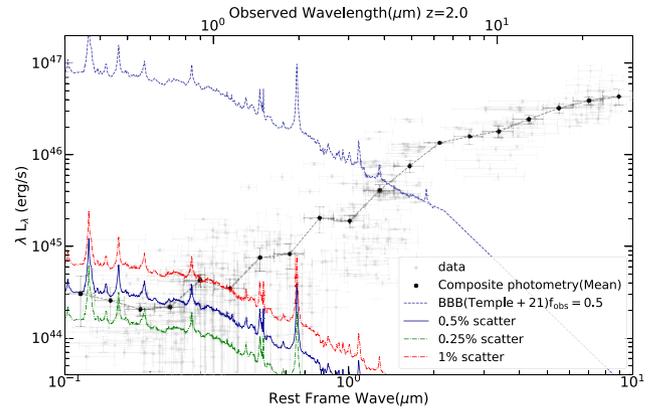

**Figure 12.** The scattered BBB fitting result. The grey data points are the scaled photometries, and the black dots are our mean composite photometry. The blue dashed line is the BBB template from Temple et al. (2021) assuming an obscured fraction of 0.5. The solid and dash–dot lines are the results with different scatter fractions. The solid blue line shows the result of applying a 0.5 per cent scatter fraction on the BBB template.

Assuming the obscuration fraction to be 0.5 (e.g. Polletta et al. 2007; Merloni et al. 2014; Lusso et al. 2015), we scaled the BBB template from Temple et al. (2021) to twice the luminosity of our mean dust torus. Then we apply various $E(B - V)$ values using the Small Magellanic Cloud reddening law to see if the reddened BBB model can fit our data. The results are shown in Fig. 11. The blue solid line is the scaled BBB template assuming an obscured fraction equals 0.5. The coloured dashed lines are the reddened BBB models. None of the reddening models fit our rest-UV data. But some models with $E(B - V) \sim 1.5$ fit our rest-optical data. The reddening BBB light may contribute to our rest-optical data but cannot explain our rest-UV light.

### 4.3 Could the UV light come from the scattered light?

Scattered emission has often been invoked to explain the polarization and the rest-UV spectral continuum in type 2 QSOs (Alexandroff et al. 2018; Zakamska & Alexandroff 2023). This scenario has also been applied to explain the 'v-shape' SED of the LRDs (Greene et al. 2024). In this scattered light scenario, the blue slope is due to the scattered light of the BBB model, and the red continuum is explained by the hot torus. We want to test if the scattered light is a possible explanation for the rest-UV light of our SEDs. To quantitatively consider a scattered BBB we proceed as follows. We take the BBB template from Temple et al. (2021) and scale it to twice the mean torus luminosity of our SED (assuming a 0.5 obscured fraction), as discussed in Section 4.2 and described by equation (7).

We assume that the BBB along our line of sight is completely extinct, but a small fraction of the light, $f_{scatter}$, is scattered into our direction. The scattering agent could be dust or electrons, but distinguishing between these two different scatterings is challenging (Zakamska et al. 2005). Here we just follow what has been done in some LRD studies (e.g. Akins et al. 2024; Greene et al. 2024): assume scattering agents are the electrons; this scattering produces a similar SED shape as the unobscured QSOs. With the assumed scattered fraction, the observed BBB should be

$$L_{BBB(observed)} = L_{BBB} \times f_{scatter} = L_{torus} \frac{f_{scatter}}{f_{obscuration}}. \qquad (8)$$

The result is shown in Fig. 12.

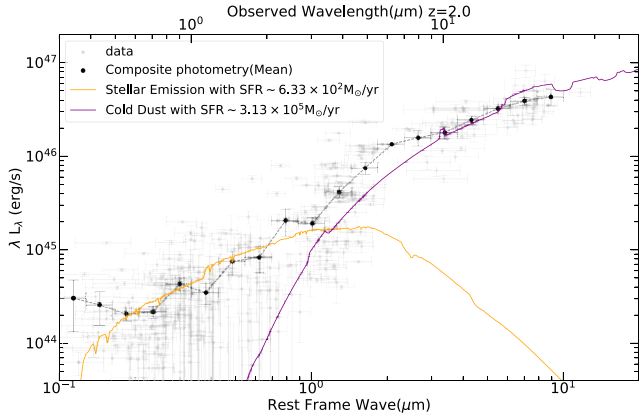

**Figure 13.** The fitting results of the SEDs when only using the stellar emission and cold dust. The grey data points are the scaled photometry, and the black dots are the mean photometry. The stellar emission and cold dust templates are shown in orange and purple. If the rest-UV and optical light are all from the star formation, then the galaxy should have a SFR $\sim 6.33 \times 10^2\,M_\odot\,yr^{-1}$. If the IR emission is not from the dust torus but the cold dust, the SFR should be $\sim 3.13 \times 10^5\,M_\odot\,yr^{-1}$. The SFR in IR seems too high for typical star-forming galaxies at $z \sim 2$ so we prefer the AGN-dominating scenario in the IR light.

The rest-UV light can be fitted by 0.5 per cent of the BBB template. Depending on the different values of the obscured fraction (0.5–1), the scattered fraction has a range of 0.5–1 per cent. This result is similar to some analysis of *JWST* LRDs (e.g. Greene et al. 2024). If the scattered light is the real contributor to the rest-UV light, then we shall see the scatter broad emission lines in the spectra. We will discuss this in a future paper (Wang, in preparation).

### 4.4 Could the SED be explained without AGN?

We also conduct the SED fitting with only the galaxy component to see if it is possible to explain these targets without an AGN component. We set the 'PRIOR_AGNfraction' and 'turn_on_AGN' both to False in the AGN fitter pipeline, and we only use the stellar emission (from Bruzual & Charlot 2003) and cold dust model (from Schreiber et al. 2018) to do the fitting. The basic assumptions of these two models are described in the previous section about SED fitting. We run the fit for every target and take the mean fitting value of the stellar emission and cold dust. The templates with corresponding SFR values are shown in Fig. 13.

The pipeline fits a star-forming galaxy to the rets-frame UV. AGNFITTER makes two independent dereddened SFR estimates in the optical/UV and IR. For the SFR in the rest-optical/UV, the pipeline fits a galaxy template and gets the total stellar mass and the age of SFH. Then the star formation is calculated by

$$\text{SFR}_{\text{opt/UV}} = 368\,M_\odot\,yr^{-1}\,10^{\text{GA}-4} \left( \frac{M_*}{10^{10}\,M_\odot} \right) \left( \frac{10^{11}}{\tau} \right)$$
$$\times \exp\left(1 - \text{age}/\tau\right), \tag{9}$$

where GA is the normalization log parameter to scale the emission of the galaxy template to the total emission needed to fit. $\tau$ is the age of the star formation history of the stellar population. $M_*$ is the mass of stellar population. If the rest-UV light comes from star formation, the SFR needs to be above $6.33 \times 10^2\,M_\odot\,yr^{-1}$ (the mean value of all the individual fittings). So for the rest-UV to optical light, the star formation is a possible explanation.

We also apply an SED fitting pipeline, MEPHISTO (a large language model-based fitting with CIGALE; see Sun et al. 2024), to the composite photometry. The fitting results give a $>0.9$ AGN fraction to the IR emission. If we turn off the AGN component in this pipeline, the fitting result in the rest-UV/optical is a galaxy model with SFR $\sim 600\,M_\odot\,yr^{-1}$. These results are consistent with AGNFITTER.

The pipeline also gives a cold dust model template fitting in the rest-frame 1–10 $\mu$m. For the SFR above 8 $\mu$m, the pipeline estimates the SFR by using

$$\text{SFR}_{\text{IR}} = 3.88 \times 10^5\,M_\odot\,yr^{-1} \left( \frac{L_{\text{IR}}}{10^{49}\,\text{erg s}^{-1}} \right), \tag{10}$$

where $L_{\text{IR}}$ is the integrated luminosity of the starburst template in 8–1000 $\mu$m. This luminosity is usually after the subtraction of the AGN hot dust torus contribution to the IR luminosity. However, in the 'no-AGN' fitting we turn off the AGN contribution and assume all the luminosities are from this starburst template. Given the shape of the cold dust template (it keeps rising above rest-frame 100 $\mu$m), the mean value of the fitted $L_{\text{IR}}$ with the cold dust template is $\sim 10^{48.9}$ erg s$^{-1}$. If the energy at this wavelength range is not caused by the hot dust torus but by the starburst, the SFR needs to be above a mean value of $3.13 \times 10^5\,M_\odot\,yr^{-1}$ and the PAH fraction 0.5 per cent. This would imply an absurdly high SFR in the IR and such galaxies would be unphysical since no galaxies have been observed to have this SFR value in the IR.

Considering the relatively small number density of these luminous objects, it is not completely impossible to find such high SFR galaxies. But considering the spectra and the SED results, we prefer the explanation that these targets are AGN-dominated sources. We conclude that the bright *WISE* W4 flux is from the hot dust torus. For the rest-UV/optical, we have three possible scenarios: (1) the rest-UV/optical light comes from the galaxy with SFR $\sim 633\,M_\odot\,yr^{-1}$; (2) the rest-UV/optical light comes from the 1 per cent scattered light of the QSO and the stellar emission from the host galaxy with $M_* \sim 10^{11}\,M_\odot$; and (3) the rest-UV light comes from the 0.5 per cent scattered light of the QSO and the rest optical light is contributed by reddening of the BBB with $E(B-V) \sim 1.5$.

### 4.5 Comparison to *JWST* LRDs

The large population of LRDs discovered by *JWST* indicates a higher obscured fraction at $z > 4$ if they are all AGNs (e.g. Matthee et al. 2024; Pizzati et al. 2024). However, their true nature is still unclear. Their 'V-shape' SED makes them seem to be a complex combination of AGNs and galaxies (e.g. Greene et al. 2024; Matthee et al. 2024). For now, the redshift distributions of the LRDs are mainly at $z = 4$–9 (e.g. Kocevski et al. 2023; Akins et al. 2024) resulting in an unknown obscured fraction at $2 < z < 4$. The luminous type 2 QSOs in our sample indicate that the number density of luminous type 2 QSOs (average $L_{\text{bol}} = 10^{47.04}$ erg s$^{-1}$, see Section 2) in the SDSS Stripe 82 region have $n_{\text{type 2}} \simeq 0.55\,\text{deg}^{-2}$, whereas type 1 QSOs of comparable $L_{\text{bol}}$ have $n_{\text{type 1}} \simeq 0.65\,\text{deg}^{-2}$ (see Section 2), which implies the obscured:unobscured ratio is approximately 1:1 at the luminosity of our sample.

Our type 2 QSOs represent the most luminous objects ($L_{\text{bol}} \sim 10^{47.04}$ erg s$^{-1}$) at $z \sim 2$, and LRDs constitute relatively faint targets ($\log L_{\text{bol}} = 45.5$; see Akins et al. 2024) at $z > 4$. We compare the photometry and SED of our sample to the composite LRD photometry and SED taken from Akins et al. (2024), to investigate the relationship between these two populations. The results are shown in Fig. 14. The composite photometry of our type 2 sample is shown as black points. We have shown the model with 0.5 per cent scattered light (see Fig. 12) in blue and $E(B - V) = 1.5$ (see Fig. 11) in purple. The dust torus fit from Fig. 10 is shown in red.

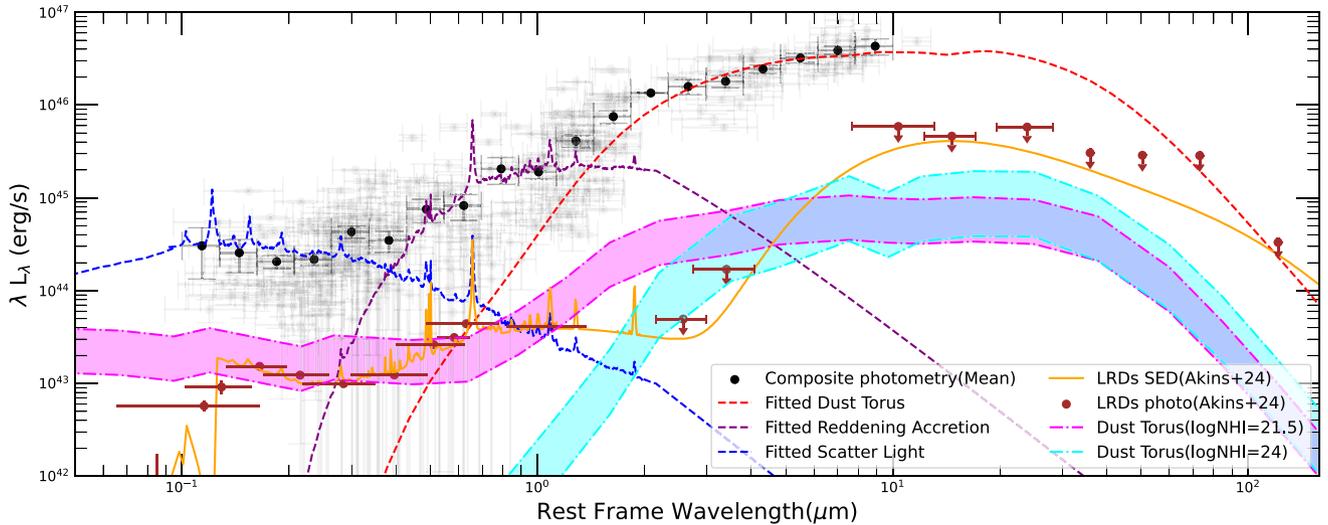

**Figure 14.** The composite photometry and SED from our type 2 QSO sample and *JWST* LRDs. The composite photometry of our type 2 sample is shown as black points. We have shown the model with 0.5 per cent scattered light (see Fig. 12) in blue and $E(B - V) = 1.5$ (see Fig. 11) in purple. The dust torus fit from Fig. 10 is shown in red. The LRD photometry from Akins et al. (2024) is shown as brown points and their SED estimate as the orange line. The SED of our type 2 QSO sample and the LRDs exhibit a similar shape below rest-frame wavelengths of 1 µm. This similarity indicates that these IR-selected type 2 QSOs may share similar UV–optical properties to LRDs. But the SED shapes appear rather different above rest-frame 1 µm. Our type 2 QSO sample has median bolometric luminosity log $L_{bol} = 47.04$ and the LRDs have median bolometric luminosity log $L_{bol} = 45.5$. Our targets are a factor of 10 brighter than the typical *JWST* LRDs. We plot two hot dust torus templates from Stalevski et al. (2016) scaled to the same bolometric luminosity as LRDs with different neutral hydrogen column densities as magenta (log $N_{H1} = 21.5$) and cyan (log $N_{H1} = 24$). These two templates were normalized by $L_{torus} = 10^{45.5}$ erg s$^{-1}$ assuming the obscured fraction, $R$, is unity (see equation 7). The bolometric luminosity estimate of LRD sample from Akins et al. (2024) assumes the light is entirely contributed by the AGN. However, given that galaxy light could also contribute, their bolometric luminosity estimate is an upper limit (the torus SED would move down). Furthermore, because the LRD sample from Akins et al. (2024) is a photometrically selected sample, potential contaminants could bias the stacked rest-frame near-IR photometry low, so a clean spectroscopic LRD sample selected via broad emission lines could yield higher near-IR stacked flux (at around rest-frame 3 µm, so the upper limits of LRDs photometry could move up). To account for both of these issues we simply allow for $L_{bol}$ normalization of the torus templates to be a factor of 3 lower (indicated by the lower set of torus lines) and thus assume that the true normalization is within the shaded region. This indicates hot dust emission from LRDs cannot currently be ruled out. The similar SED shapes of our type 2 QSOs and LRDs, along with the detection of broad Balmer's emission lines in some of the type 2 QSOs, suggests that the LRDs may be higher $z$ bolometrically fainter analogues of our type 2 QSOs.

The LRD photometry from Akins et al. (2024) is shown as brown points and their SED estimate as the orange line. The SED of our type 2 QSO sample and the LRDs exhibit a similar shape below rest-frame wavelengths of 1 µm. This similarity indicates that these IR-selected type 2 QSOs may share similar UV–optical properties to LRDs. But the SED shapes appear rather different above rest-frame 1 µm.

This difference is illustrated in Fig. 14, where one observes that the SEDs of LRDs and type 2 QSOs seem to have different shapes in the torus region. In our composite SED, the SED above rest-frame 3 µm is dominated by the dust torus. The hot dust torus fit to our targets is shown as the red dashed line, this is the mean fit of all the individual fits (same as the red solid line in Fig. 10). To investigate potential hot dust emission from LRDs, we plot two hot dust torus templates from Stalevski et al. (2016) scaled to the same bolometric luminosity as LRDs with different neutral hydrogen column densities as magenta (log $N_{H1} = 21.5$) and cyan (log $N_{H1} = 24$). Specifically, these templates were normalized by $L_{torus} = 10^{45.5}$ erg s$^{-1}$ assuming the obscured fraction, $R$, is unity (see equation 7). The bolometric luminosity estimate of LRD sample from Akins et al. (2024) is using a scattered light + power-law fit, assuming the light is entirely contributed by the AGN. However, given that galaxy light could also contribute, their bolometric luminosity estimate is an upper limit (the torus SED would move down). Furthermore, because the LRD sample from Akins et al. (2024) is a photometrically selected sample, potential contaminants (i.e. brown dwarfs or objects that do not have broad lines) could bias the stacked rest-frame near-IR photometry low, so a clean spectroscopic LRD sample (selected via broad emission lines) could yield higher near-IR stacked flux (at around rest-frame 3 µm, so the upper limits of LRDs photometry could move up). To account for both of these issues we simply allow for $L_{bol}$ normalization of the torus templates to be a factor of 3 lower (indicated by the lower set of torus lines) and thus assume that the true normalization is within the shaded regions. Given these caveats, Fig. 14 indicates hot dust emission from LRDs cannot currently be ruled out. The similar SED shapes of our type 2 QSOs and LRDs, along with the detection of broad Balmer's emission lines in some of the type 2 QSOs, suggests that the LRDs may be higher $z$ bolometrically fainter analogues of our type 2 QSOs.

Our selection was designed to find luminous type 2 QSOs at $z > 2$ and the sample in this paper is a pilot study in the SDSS Stripe 82 region. However, our targets have spectra and photometry that resemble LRDs. Obtaining high-quality spectra of more type 2 QSOs, particularly at cosmic times coeval with LRDs, may help us better understand if *JWST* LRD can have a hot dust torus, and may also answer the question of why LRDs are missing at $z < 4$.

## 5 CONCLUSION

In this paper, we present the mid-IR selection, SED fitting, and composite photometry for 23 spectroscopically confirmed type 2 QSOs at $z \sim 2$ in the SDSS Stripe 82 region. The main results are as follows.

(i) These targets are selected to be mid-IR bright (flux$_{W4}$ > 5 mJy, i.e. $12.62 < W4 < 14.62$ ABmag), optically faint ($r > 23$), or with red colour ($r - W4 > 8.38$). After spectroscopic observations using Gemini North/GNIRS and Keck/LRIS, 23 targets are successfully identified to be type 2 QSOs at $z = 0.88$–3.49, yielding a high success rate (>96 per cent). Among the 23 identified targets, 12 are above $z > 2$. The mid-IR selection is an efficient way to find luminous type 2 QSOs at redshifts $z > 2$.

(ii) By collecting photometry from optical to IR, we conduct SED fitting to these targets. They are the most IR-luminous QSOs and have bolometric luminosity at $L_{\rm bol} \sim 10^{46.28}$–$10^{48.08}$ erg s$^{-1}$. These targets all have hot dust torus components and high AGN fractions, proving they are AGN-dominated objects. If they do not have AGN components, then the SFR of the galaxy should be >$3.13 \times 10^5$ M$_\odot$ yr$^{-1}$ to produce the *WISE* W4 flux. This value is too high for the typical $z \sim 2$ galaxies.

(iii) We have investigated the components in the composite SED. The IR emission is dominated by a dust torus with a mean value $L_{\rm torus} = 10^{46.84}$ erg s$^{-1}$. Assuming the obscured fraction equal to 0.5, the mean value of the bolometric luminosity of our targets is $L_{\rm bol} = 10^{47.04}$ erg s$^{-1}$. For the rest-frame UV to optical, it could be dominated by: (a) a 0.5–1 per cent contribution of scattered light from the accretion disc and stellar emission from a galaxy with $M_* \sim 10^{11}$ M$_\odot$; (b) a 0.5–1 per cent contribution of scattered light from the accretion disc and reddened accretion disc (BBB) with $E(B - V) \sim 1.5$; and (c) a star-forming galaxy with SFR $\sim 633$ M$_\odot$ yr$^{-1}$ and $M_* \sim 10^{10.2}$ M$_\odot$.

(iv) Composite photometry for type 2 QSOs is generated by normalizing the photometry using the SED fitting results. This composite photometry and SED model will guide us in finding these obscured QSOs at higher redshift in surveys like *Euclid*. Compared to previously published type 2 SEDs, our new composite SED has a higher bolometric luminosity and redder optical–IR colour.

(v) We compare our composite photometry and SED results to the stacked photometry and SED of LRDs. The two different samples exhibit a similar SED shape at different luminosity levels. Our type 2 QSOs have a mean bolometric luminosity at $L_{\rm bol} = 10^{47.04}$ erg s$^{-1}$ and the mean bolometric luminosity of LRDs from Akins et al. (2024) is $L_{\rm bol} = 10^{45.5}$ erg s$^{-1}$. By putting different hot dust torus templates on the LRDs' SED, we cannot conclude that LRDs do not have hot dust. Considering the similar SED shape, LRDs with hot dust could be drawn from the fainter type 2 QSO population at higher redshifts.

Finding more luminous type 2 QSOs at high redshift will be the key to understanding the obscured fraction across cosmic time and the evolution of AGNs. Mid-IR selection is an efficient and unbiased way to isolate these targets, and their spectra show huge diversity in the rest-frame UV and optical. We will publish the spectra of these targets and the results of spectroscopic analysis in an upcoming paper.

## ACKNOWLEDGEMENTS

BW and ZC were supported by the National Key Research and Development Program of China (grant no. 2023YFA1605600), the National Science Foundation of China (grant no. 12073014), the science research grants from the China Manned Space Project with no. CMS-CSST-2021-A05, and Tsinghua Initiative Scientific Research Program (no. 20223080023).

J-TS was supported by the Deutsche Forschungsgemeinschaft (DFG, German Research Foundation) – project number 518006966.

We acknowledge the fruitful discussion with Manda Banerji and Roberto Maiolino.

## DATA AVAILABILITY

The photometries used in this paper are all public data from DESI Legacy Survey, PS1, UKIDSS, *Spitzer*, and *WISE* (see Tables A1 and A2). The Gemini/GNIRS spectra are public and easy to get on the GNIRS data archive under the proposal ID GN-2017B-Q-51. The Keck/LRIS spectra will be published in our next paper.

# APPENDIX A: PHOTOMETRY FLUX AND CUT-OUTS

In this section, we show the photometry in flux (μJy) units. The results are shown in Tables A1 and A2.

**Table A1.** The target name, redshift, and flux (μJy) in each band of all 24 targets.

| Target | Redshift | g | r | i | z | y | Y | J | H | K |
|---|---|---|---|---|---|---|---|---|---|---|
| J0024−0012 | 1.53 | 0.920 ± 0.083 | 1.415 ± 0.073 | 1.086 ± 0.337 | 3.930 ± 0.237 | 1.674 ± 1.418 | 2.035 ± 3.012 | 8.983 ± 3.914 | 9.816 ± 5.111 | 4.978 ± 5.300 |
| J0041−0029 | 2.09 | 0.837 ± 0.086 | 1.136 ± 0.115 | −0.081 ± 0.300 | 2.814 ± 0.334 | −0.970 ± 1.717 | 0.000 ± 0.000 | −2.817 ± 4.230 | −10.66 ± 6.336 | 7.559 ± 5.897 |
| J0047−0003 | – | 0.241 ± 0.127 | 0.862 ± 0.178 | −0.580 ± 0.330 | 2.301 ± 0.549 | −2.210 ± 1.541 | −1.472 ± 1.974 | 1.941 ± 2.971 | 1.850 ± 3.415 | 3.801 ± 4.469 |
| J0054−0047 | 2.17 | 0.232 ± 0.091 | 0.483 ± 0.108 | 0.345 ± 0.316 | 3.118 ± 0.323 | 3.577 ± 1.383 | 2.191 ± 2.307 | 3.589 ± 2.628 | 4.974 ± 5.036 | 5.680 ± 4.296 |
| J0105−0023 | 1.87 | 0.487 ± 0.057 | 0.893 ± 0.076 | 1.021 ± 0.359 | 1.889 ± 0.201 | −1.159 ± 1.989 | 0.988 ± 2.547 | 0.973 ± 3.303 | 2.436 ± 4.593 | 2.456 ± 5.666 |
| J0112−0016 | 2.99 | 0.249 ± 0.069 | 0.525 ± 0.106 | 0.570 ± 0.322 | 0.594 ± 0.204 | 1.022 ± 1.446 | 0.833 ± 2.440 | 2.051 ± 3.083 | 1.387 ± 5.103 | −3.471 ± 5.062 |
| J0113−0029 | 2.33 | 0.548 ± 0.073 | 0.833 ± 0.100 | 0.366 ± 0.310 | 1.496 ± 0.283 | 2.911 ± 1.677 | 2.062 ± 2.195 | −0.820 ± 2.741 | −2.659 ± 5.037 | 2.063 ± 5.059 |
| J0130−0009 | 2.50 | 0.556 ± 0.121 | 0.217 ± 0.183 | 0.246 ± 0.301 | −0.092 ± 0.484 | 2.807 ± 1.594 | −3.672 ± 3.411 | 1.270 ± 4.908 | 3.628 ± 4.519 | 8.048 ± 5.375 |
| J0149−0052 | 1.85 | 0.608 ± 0.053 | 0.913 ± 0.068 | 0.516 ± 0.240 | 1.722 ± 0.201 | −1.309 ± 1.479 | 3.521 ± 2.786 | 8.330 ± 3.948 | 11.80 ± 5.336 | 2.384 ± 5.565 |
| J0150+0056 | 3.49 | 0.892 ± 0.060 | 1.367 ± 0.076 | −3.120 ± 0.241 | 1.364 ± 0.211 | −7.570 ± 1.667 | 1.065 ± 2.924 | −9.321 ± 3.864 | 1.166 ± 5.299 | −1.912 ± 5.304 |
| J0152−0024 | 2.78 | 1.703 ± 0.069 | 2.671 ± 0.103 | 0.874 ± 0.295 | 3.154 ± 0.212 | 1.216 ± 1.618 | 5.462 ± 5.086 | 4.177 ± 4.011 | 5.624 ± 3.977 | 14.68 ± 4.263 |
| J0213−0024 | 1.81 | 1.338 ± 0.069 | 2.039 ± 0.084 | 2.290 ± 0.317 | 12.350 ± 0.266 | 8.604 ± 1.656 | 12.75 ± 2.107 | 39.90 ± 2.783 | 67.51 ± 5.583 | 121.6 ± 4.993 |
| J0214−0000 | 1.63 | 1.925 ± 0.092 | 2.587 ± 0.144 | 1.205 ± 0.302 | 5.301 ± 0.423 | 1.062 ± 1.662 | 1.790 ± 1.992 | 7.353 ± 2.914 | 5.522 ± 5.623 | 13.11 ± 4.839 |
| J0215+0042 | 0.88 | 0.975 ± 0.066 | 2.797 ± 0.088 | 2.832 ± 0.353 | 11.138 ± 0.285 | 9.327 ± 1.853 | 10.21 ± 2.896 | 0.000 ± 0.000 | 28.54 ± 5.478 | 74.22 ± 4.763 |
| J0221+0050 | 2.48 | 0.538 ± 0.103 | 0.605 ± 0.154 | 0.367 ± 0.345 | 1.241 ± 0.453 | −2.423 ± 1.384 | 6.121 ± 2.896 | −5.019 ± 3.769 | −0.086 ± 5.384 | 12.34 ± 4.788 |
| J2229+0022 | 1.93 | 0.772 ± 0.105 | 1.796 ± 0.132 | 0.961 ± 0.280 | 4.107 ± 0.323 | 3.533 ± 1.915 | 4.474 ± 2.437 | 8.045 ± 3.310 | 7.992 ± 4.836 | 14.77 ± 5.118 |
| J2233−0004 | 1.60 | 0.465 ± 0.093 | 1.556 ± 0.112 | 1.365 ± 0.231 | 5.770 ± 0.297 | 6.021 ± 2.217 | 2.506 ± 2.804 | 9.011 ± 3.817 | 9.618 ± 5.568 | 20.46 ± 5.639 |
| J2239−0030 | 1.91 | 0.886 ± 0.064 | 1.545 ± 0.075 | 0.650 ± 0.219 | 2.503 ± 0.204 | 0.938 ± 2.025 | −0.554 ± 4.252 | −2.106 ± 3.310 | 2.700 ± 5.627 | 12.18 ± 5.289 |
| J2239−0054 | 2.09 | 0.898 ± 0.082 | 1.235 ± 0.105 | 0.586 ± 0.243 | 3.597 ± 0.311 | 2.761 ± 2.144 | 2.966 ± 2.114 | −0.129 ± 2.943 | 5.302 ± 5.286 | 4.335 ± 4.935 |
| J2243+0017 | 1.91 | 0.996 ± 0.075 | 1.868 ± 0.085 | 1.346 ± 0.238 | 7.922 ± 0.266 | 3.336 ± 1.870 | 1.477 ± 2.489 | 18.95 ± 3.695 | 42.98 ± 6.371 | 98.60 ± 5.861 |
| J2258−0022 | 2.42 | 0.686 ± 0.071 | 0.943 ± 0.086 | 0.098 ± 0.314 | 1.981 ± 0.228 | −1.206 ± 1.909 | 1.087 ± 1.901 | −2.945 ± 2.362 | −4.417 ± 3.543 | −1.215 ± 4.506 |
| J2259−0009 | 1.89 | 1.347 ± 0.092 | 1.868 ± 0.115 | 1.486 ± 0.334 | 4.522 ± 0.325 | 0.690 ± 1.758 | 3.506 ± 1.767 | 4.562 ± 2.046 | 10.23 ± 3.541 | 25.41 ± 4.256 |
| J2329+0020 | 2.67 | 0.722 ± 0.086 | 0.988 ± 0.103 | 0.663 ± 0.340 | 2.470 ± 0.283 | 2.383 ± 1.435 | 0.411 ± 1.789 | −4.763 ± 2.653 | 12.41 ± 4.941 | 1.621 ± 4.876 |
| J2334+0031 | 2.10 | 2.293 ± 0.065 | 2.461 ± 0.080 | 2.111 ± 0.317 | 4.196 ± 0.207 | 0.574 ± 1.709 | 5.384 ± 2.215 | 1.179 ± 3.111 | 3.565 ± 6.014 | 5.487 ± 6.167 |

**Table A2.** The target name, redshift, and flux (μJy) in each band of all 24 targets.

| Target | Redshift | W1 | IRAC1 | IRAC2 | W2 | W3 | W4 | Group |
|--------|----------|-----|--------|--------|-----|-----|-----|-------|
| J0024−0012 | 1.53 | 93.46 ± 6.138 | 152.8 ± 2.081 | 386.7 ± 2.863 | 386.8 ± 15.75 | 1997 ± 152.7 | 5488 ± 1061 | Type 2 |
| J0041−0029 | 2.09 | 40.25 ± 5.652 | 33.24 ± 1.730 | 67.43 ± 1.901 | 111.2 ± 11.61 | 1101 ± 130.9 | 7056 ± 1071 | Type 2 |
| J0047+0003 | – | 32.51 ± 5.710 | 86.97 ± 18.65 | 92.68 ± 16.09 | 87.37 ± 13.88 | 2932 ± 199.8 | 10819 ± 1286 | Unknown |
| J0054+0047 | 2.17 | 51.28 ± 5.324 | 59.84 ± 1.866 | 135.6 ± 2.297 | 109.5 ± 14.80 | 1557 ± 179.4 | 6376 ± 914.8 | Type 2 |
| J0105−0023 | 1.87 | 27.09 ± 5.090 | 38.22 ± 3.964 | 133.1 ± 5.004 | 149.9 ± 11.74 | 3060 ± 168.4 | 6742 ± 1223 | Type 2 |
| J0112−0016 | 2.99 | 19.02 ± 5.225 | 26.99 ± 9.759 | 45.32 ± 6.530 | 29.48 ± 10.52 | 832.0 ± 144.2 | 6010 ± 1076 | Type 2 |
| J0113+0029 | 2.33 | 14.49 ± 5.303 | 27.03 ± 5.073 | 79.15 ± 4.516 | 50.85 ± 11.12 | 1585 ± 138.0 | 5331 ± 998.4 | Type 2 |
| J0130+0009 | 2.50 | 50.89 ± 4.492 | 78.92 ± 4.589 | 176.9 ± 4.799 | 176.4 ± 10.99 | 1942 ± 117.3 | 5488 ± 878.2 | Type 2 |
| J0149+0052 | 1.85 | 93.65 ± 5.306 | 131.5 ± 2.094 | 300.8 ± 2.784 | 327.3 ± 14.43 | 2144 ± 134.7 | 5331 ± 1019 | Type 2 |
| J0150+0056 | 3.49 | 90.75 ± 5.093 | 35.57 ± 1.954 | 37.48 ± 1.905 | 89.26 ± 10.07 | 1380 ± 109.2 | 10349 ± 852.1 | Type 2 |
| J0152−0024 | 2.78 | 125.78 ± 5.772 | 115.38 ± 2.202 | 215.87 ± 2.351 | 246.7 ± 12.13 | 2364 ± 134.3 | 5958 ± 857.3 | Type 2 |
| J0213+0024 | 1.81 | 354.1 ± 10.17 | 459.2 ± 3.010 | 790.3 ± 3.914 | 760.1 ± 21.84 | 3170 ± 145.7 | 7840 ± 873.0 | Redden type 1 |
| J0214−0000 | 1.63 | 54.76 ± 4.531 | 38.17 ± 3.128 | 172.6 ± 2.417 | 152.3 ± 10.82 | 2327 ± 127.9 | 6167 ± 768.4 | Type 2 |
| J0215+0042 | 0.88 | 311.5 ± 9.586 | 418.9 ± 2.972 | 769.2 ± 3.870 | 754.6 ± 21.79 | 3408 ± 142.2 | 9199 ± 977.5 | Type 2 |
| J0221+0050 | 2.48 | 107.6 ± 5.636 | 101.7 ± 2.178 | 188.6 ± 2.462 | 231.0 ± 11.42 | 1713 ± 124.1 | 6899 ± 888.6 | Type 2 |
| J2229+0022 | 1.93 | 126.6 ± 6.603 | 58.10 ± 1.855 | 78.46 ± 2.017 | 165.2 ± 12.45 | 3280 ± 173.7 | 12649 ± 1280 | Type 2 |
| J2233−0004 | 1.60 | 52.63 ± 5.575 | 61.97 ± 1.988 | 92.77 ± 1.922 | 100.9 ± 11.55 | 868.7 ± 147.7 | 6220 ± 1076 | Type 2 |
| J2239−0030 | 1.91 | 31.73 ± 5.284 | 27.25 ± 1.664 | 83.54 ± 1.807 | 78.17 ± 11.77 | 2895 ± 148.8 | 8572 ± 1223 | Type 2 |
| J2239−0054 | 2.09 | 62.31 ± 5.788 | 50.66 ± 1.987 | 129.6 ± 3.700 | 150.7 ± 12.26 | 2089 ± 156.7 | 6585 ± 1113 | Type 2 |
| J2243+0017 | 1.91 | 313.5 ± 10.42 | 360.0 ± 2.803 | 544.6 ± 3.385 | 516.6 ± 18.92 | 2620 ± 192.4 | 8206 ± 1312 | Type 2 |
| J2258−0022 | 2.42 | 11.28 ± 5.206 | 33.67 ± 1.593 | 80.27 ± 2.038 | 78.98 ± 11.82 | 938.3 ± 136.3 | 5801 ± 1050 | Type 2 |
| J2259−0009 | 1.89 | 116.3 ± 6.912 | 161.8 ± 2.190 | 346.9 ± 2.910 | 346.2 ± 15.40 | 2217 ± 173.9 | 6533 ± 1108 | Type 2 |
| J2329+0020 | 2.67 | 9.38 ± 4.973 | 25.84 ± 1.796 | 37.93 ± 1.502 | 26.72 ± 10.63 | 1057 ± 143.5 | 5854 ± 1019 | Type 2 |
| J2334+0021 | 2.10 | 65.21 ± 5.402 | 63.12 ± 2.001 | 138.5 ± 2.252 | 150.7 ± 12.99 | 1526 ± 140.7 | 7369 ± 1102 | Type 2 |

## APPENDIX B: SED FITTING RESULTS

The output parameters for every target are shown in Tables B1 and B2. At the end of Table B2, we show the bolometric luminosity results both from our SED fitting results and the results from Ishikawa et al. (2023). In Ishikawa et al. (2023), the bolometric luminosity is calculated by using the IR-bolometric correction:

$$L_{bol} = 8 \times L_{3.45\,\mu m}. \qquad (B1)$$

The SED fitting results and the cut-outs for all targets are shown in Fig. B1. The SED fitting provides one possible explanation.

For four targets (J0152−0024, J0221+0050, J2243+0017, and J2259−0009), we have shown two fitting results in Fig. B2: blue QSO with red galaxy, and red QSO with blue galaxy. In the second fitting, if the stellar emission from the galaxy contributes to the 1 μm, then the stellar mass should be above $10^{12}\,M_\odot$. This mass is too high for a typical $z \sim 2$ galaxy. Therefore, the blue galaxy and red QSO components are preferred. For the rest targets, a galaxy with around or below $\sim 10^{11}\,M_\odot$ is reasonable (see Section 4.1), so the fitting here provides one possible scenario.

**Table B1.** The SED fitting parameter results for the 23 identified targets.

| Target | Redshift | Z | τ | Age | E(B−V)_bal | Tdust | fracPAH | incl | E(B−V)_bbb | GA | SB | TO | BB | logMstar | SFR_opt | LIR (8–1000) | Lbb (0.1–1) | Lbbdered (0.1–1) | Lga (0.1–1) |
|---|---|---|---|---|---|---|---|---|---|---|---|---|---|---|---|---|---|---|---|
| J0024−0012 | 1.53 | 1.6810 | 6.9800 | 7.8182 | 0.7120 | 27.2347 | 0.0303 | 8.1523 | 0.2082 | 4.2307 | 1.6316 | 5.3466 | 1.0776 | 10.5229 | 434.9949 | 42.8459 | 44.3679 | 45.1110 | 44.4932 |
| J0041−0029 | 2.09 | 1.4472 | 4.0753 | 10.0249 | 0.6646 | 28.4687 | 0.0023 | 31.7455 | 0.1089 | 5.0961 | 6.7688 | 1.6320 | 0.9814 | 11.4180 | 50.1857 | 43.5091 | 44.6566 | 45.1656 | 44.6730 |
| J0054−0047 | 2.17 | 1.5840 | 3.7162 | 10.7039 | 0.7300 | 27.9411 | 0.0265 | 66.0530 | 0.3622 | 5.7174 | 1.5443 | 5.7409 | 1.6073 | 11.2123 | 50.6034 | 43.1232 | 44.8277 | 45.8135 | 44.7850 |
| J0105−0023 | 1.87 | 1.9576 | 11.8830 | 11.8387 | 0.7397 | 41.4934 | 0.0522 | 54.6817 | 0.7351 | 4.6657 | 2.1924 | 5.8207 | 1.5676 | 10.8731 | 507.0649 | 42.4563 | 44.3831 | 45.3726 | 44.5066 |
| J0112−0016 | 2.99 | 1.1911 | 5.8721 | 8.2299 | 0.3771 | 29.4897 | 0.0240 | 49.1667 | 0.2378 | 3.4756 | 0.5825 | 5.7752 | 1.0068 | 9.8624 | 32.5165 | 43.1620 | 44.6795 | 45.3104 | 44.9160 |
| J0113−0029 | 2.33 | 1.1536 | 5.7306 | 10.3696 | 0.4847 | 29.3178 | 0.0248 | 49.6961 | 0.0873 | 4.0653 | 0.1013 | 5.6437 | 0.7424 | 10.5536 | 5.1561 | 42.4983 | 44.5668 | 44.9839 | 44.3058 |
| J0130+0009 | 2.50 | 1.6263 | 5.4154 | 10.7092 | 0.7023 | 28.9261 | 0.0263 | 21.2069 | 0.0467 | 5.8507 | 1.6675 | 5.5199 | 0.0966 | 11.4590 | 189.2402 | 43.4269 | 44.3066 | 44.7795 | 45.1120 |
| J0149+0052 | 1.85 | 1.6948 | 4.7033 | 10.7209 | 0.7641 | 28.3501 | 0.0266 | 10.2024 | 0.0846 | 5.8213 | 1.7773 | 5.3962 | 0.6961 | 11.1722 | 106.2134 | 43.1881 | 44.3224 | 44.7361 | 44.7904 |
| J0150+0056 | 3.49 | 1.6948 | 1.0585 | 11.0325 | 0.6085 | 28.4493 | 0.0282 | 86.8068 | 0.0044 | 5.9278 | 1.7363 | 5.6751 | 0.4565 | 11.2209 | 0.1359 | 43.1298 | 44.2658 | 44.3259 | 44.5883 |
| J0152−0024 | 2.78 | 0.5352 | 1.7751 | 10.8364 | 0.7067 | 27.2416 | 0.0266 | 39.6934 | 2.74 | 6.4626 | 2.2052 | 5.7474 | 1.0648 | 11.1667 | 93.7514 | 43.8365 | 45.1078 | 45.3097 | 44.5180 |
| J0213−0024 | 1.81 | 1.6917 | 1.6061 | 10.0564 | 0.3237 | 28.8887 | 0.0274 | 10.1561 | 2.4665 | 6.3308 | 2.3163 | 5.5380 | 2.3618 | 10.8450 | 46.5527 | 43.7895 | 45.1820 | 46.3227 | 45.4304 |
| J0214−0000 | 1.63 | 1.2426 | 6.0328 | 9.7875 | 0.3096 | 30.1234 | 0.0199 | 50.0929 | 0.0370 | 4.9167 | 1.2459 | 5.5834 | 0.8222 | 11.2322 | 29.8328 | 42.7431 | 44.5603 | 44.7767 | 44.6240 |
| J0215+0042 | 0.88 | 1.9602 | 3.8386 | 12.6332 | 0.3498 | 41.4789 | 0.0525 | 31.6626 | 0.6686 | 5.8134 | 3.1118 | 5.4573 | 1.9296 | 11.5640 | 11.0408 | 42.7527 | 43.9014 | 45.0687 | 43.3064 |
| J0221+0050 | 2.48 | 1.5762 | 1.8106 | 10.9809 | 0.7428 | 27.8979 | 0.0285 | 76.0196 | 2.7857 | 6.4285 | 2.1325 | 5.9071 | 0.3939 | 11.0817 | 37.6508 | 43.7388 | 44.5719 | 44.9379 | 45.3175 |
| J2229+0022 | 1.93 | 1.7219 | 6.3037 | 7.8065 | 0.7460 | 29.2671 | 0.0326 | 86.4411 | 0.0840 | 4.7518 | 1.9310 | 6.0916 | 6.6356 | 11.1440 | 2839.6976 | 43.4570 | 44.6515 | 45.2699 | 45.0469 |
| J2233−0004 | 1.60 | 1.2897 | 6.0352 | 7.6000 | 0.5403 | 29.7165 | 0.0254 | 46.0731 | 0.6806 | 3.8647 | 1.1329 | 5.1657 | 2.1742 | 10.0568 | 189.7350 | 42.7734 | 44.7498 | 46.1868 | 44.5624 |
| J2239−0030 | 1.91 | 1.4208 | 7.0799 | 9.5515 | 0.3718 | 28.6576 | 0.0250 | 68.1012 | 0.0826 | 4.6808 | 1.564 | 5.9170 | 0.0425 | 11.0331 | 62.3442 | 42.7324 | 44.5061 | 44.8952 | 44.6788 |
| J2239−0054 | 2.09 | 1.5676 | 5.0256 | 10.9326 | 0.7113 | 28.0132 | 0.0262 | 49.2779 | 0.0963 | 5.7061 | 1.6254 | 5.6671 | 0.9402 | 11.1217 | 116.9086 | 43.1890 | 44.6084 | 45.0901 | 44.8606 |
| J2243+0017 | 1.91 | 1.6750 | 0.8986 | 9.7223 | 0.2782 | 32.3535 | 0.0193 | 38.7865 | 2.0232 | 6.1712 | 2.0092 | 5.6191 | 1.7473 | 10.8846 | 19.2800 | 43.8263 | 44.8796 | 45.8133 | 45.4963 |
| J2258−0022 | 2.42 | 1.9114 | 11.8719 | 11.8695 | 0.7374 | 41.2849 | 0.0528 | 89.8301 | 0.6080 | 3.6661 | 1.1325 | 5.8867 | 2.9024 | 10.1802 | 12.7960 | 41.5736 | 45.7385 | 46.9520 | 44.0053 |
| J2259−0090 | 1.89 | 1.6750 | 6.8408 | 9.5089 | 0.7530 | 30.0158 | 0.0311 | 13.4715 | 2.9628 | 5.6052 | 1.7946 | 5.4292 | 0.9783 | 0.87 | 214.8852 | 43.3422 | 44.6893 | 45.1230 | 44.9327 |
| J2329+0020 | 2.67 | 1.0894 | 6.7406 | 9.7876 | 0.5325 | 28.6188 | 0.0236 | 70.7292 | 0.9962 | 4.0926 | 0.3886 | 5.8198 | 0.9589 | 10.8983 | 18.1007 | 42.7660 | 44.8563 | 45.3459 | 44.6916 |
| J2334+0031 | 2.10 | 1.9819 | 11.7958 | 12.1464 | 0.7979 | 41.1202 | 0.0526 | 89.4537 | 0.0318 | 6.1703 | 3.2659 | 5.8845 | 1.0363 | 11.6208 | 265.5988 | 43.3068 | 44.8717 | 44.9857 | 44.9093 |

**Table B2.** The SED fitting parameter results for the 23 identified targets. The $L_{bol}$(SED) is estimated from the $L_{bolSED}$, assuming obscured fraction equal 0.5 using equation (7). The $L_{bol}$(IR) is from Ishikawa et al. (2023).

| Target | AGNfrac (0.1-1) | Ltor (1-30) | LAGN (0.1-30) | AGNfrac (8-1000) | Lgal (0.4-0.5) | Lbbb (0.4-0.5) | AGNfrac (0.4-0.5) | Ltor(6) | Lsb (1-30) | SFR_IR | log_like | log z | $L_{bol}$ (erg s$^{-1}$) (SED) | $L_{bol}$ (erg s$^{-1}$) (IR) |
|---|---|---|---|---|---|---|---|---|---|---|---|---|---|---|
| J0024−0012 | 0.6744 | 46.4391 | 46.4430 | 0.9999 | 43.5832 | 43.5427 | 0.8322 | 46.0191 | 42.1196 | 0.2722 | −33.8878 | −62.6785 | 46.81 | 46.97 |
| J0041−0029 | 0.8724 | 46.4827 | 46.4880 | 0.9949 | 43.7262 | 43.7414 | 0.9309 | 46.0706 | 47.4496 | 1.2528 | −9.3563 | −31.8166 | 47.02 | 46.71 |
| J0054+0047 | 0.9072 | 46.8522 | 46.8558 | 1.0000 | 43.7504 | 44.0536 | 0.9579 | 46.4988 | 42.4068 | 0.5157 | −13.7694 | −40.3420 | 47.04 | 47.31 |
| J0105−0023 | 0.9799 | 46.8323 | 46.8339 | 1.0000 | 43.6339 | 43.5560 | 0.9826 | 46.4511 | 41.6117 | 0.1110 | −6.0941 | −29.7505 | 47.01 | 46.81 |
| J0112−0016 | 0.9321 | 47.2524 | 47.2533 | 1.0000 | 43.9277 | 43.8209 | 0.9471 | 46.8977 | 42.3193 | 0.5636 | −3.7301 | −26.4042 | 47.35 | 47.72 |
| J0113+0029 | 0.9982 | 46.8908 | 46.8928 | 1.0000 | 43.3088 | 43.6010 | 0.9979 | 46.5306 | 41.7199 | 0.1223 | −5.5113 | −27.9858 | 47.21 | 46.83 |
| J0130+0009 | 0.2169 | 46.9930 | 46.9934 | 0.9998 | 44.1100 | 43.2356 | 0.3237 | 46.6063 | 42.6638 | 1.0370 | −10.5551 | −36.5531 | 47.22 | 47.23 |
| J0149+0052 | 0.4147 | 46.6347 | 46.6376 | 0.9997 | 43.7737 | 43.3563 | 0.6826 | 46.2145 | 42.4162 | 0.5993 | −19.5261 | −49.4636 | 46.93 | 46.95 |
| J0150+0056 | 0.4348 | 47.04 | 46.4972 | 0.9997 | 43.3969 | 43.0866 | 0.5695 | 46.1402 | 42.3399 | 0.5232 | −246.7027 | −281.4454 | 48.08 | 46.91 |
| J0152−0024 | 0.3316 | 47.2287 | 47.2322 | 0.9996 | 44.3683 | 44.0285 | 0.4718 | 46.8550 | 43.0475 | 2.6626 | −33.5892 | −65.6274 | 47.49 | 47.35 |
| J0213+0024 | 0.4303 | 46.7481 | 46.7619 | 0.9990 | 44.3858 | 44.4118 | 0.6192 | 46.3303 | 43.0193 | 2.3945 | −137.5030 | −176.4856 | 47.04 | 47.02 |
| J0214−0000 | 0.7632 | 46.5467 | 46.5505 | 0.9999 | 43.7280 | 43.4791 | 0.7555 | 46.1730 | 41.9729 | 0.2155 | −34.4414 | −62.9592 | 46.85 | 47.03 |
| J0215+0042 | 0.3710 | 45.9643 | 45.9681 | 0.9996 | 43.3692 | 43.1317 | 0.5084 | 45.5523 | 42.0036 | 0.2196 | −26.1809 | −60.0343 | 46.28 | 46.27 |
| J0221+0050 | 0.1985 | 47.0711 | 47.0719 | 0.9996 | 44.1569 | 43.5598 | 0.3723 | 46.7176 | 42.9821 | 2.1266 | −10.5142 | −39.7358 | 47.24 | 46.25 |
| J2229+0022 | 0.3977 | 46.9937 | 46.9944 | 0.9998 | 44.0840 | 43.7725 | 0.6148 | 46.6390 | 42.7221 | 1.1114 | −31.6876 | −63.3015 | 47.32 | 47.19 |
| J2233−0004 | 0.8027 | 46.1757 | 46.1957 | 0.9999 | 43.6612 | 43.9429 | 0.8589 | 45.8034 | 42.0187 | 0.2303 | −39.6210 | −65.1663 | 46.44 | 46.23 |
| J2239−0030 | 0.9786 | 46.8828 | 46.8843 | 1.0000 | 43.8032 | 43.5241 | 0.9741 | 46.5264 | 42.0610 | 0.2097 | −23.6197 | −52.2173 | 47.20 | 47.2 |
| J2239−0054 | 0.6402 | 46.8427 | 46.8450 | 0.9999 | 43.8692 | 43.6842 | 0.7706 | 46.4884 | 42.4160 | 0.5997 | −24.0922 | −51.6975 | 47.12 | 47.12 |
| J2243+0017 | 0.2370 | 46.7894 | 46.7956 | 0.9992 | 44.5377 | 44.1003 | 0.4110 | 46.4252 | 43.0905 | 2.6017 | −79.1768 | −116.1334 | 46.94 | 47.10 |
| J2258−0022 | 1.0000 | 46.9188 | 46.9584 | 1.0000 | 43.0192 | 44.9629 | 1.0000 | 46.5653 | 40.8348 | 0.0145 | −11811.6149 | −11843.8291 | 47.01 | 46.95 |
| J2259−0010 | 0.4935 | 46.6681 | 46.6728 | 0.9996 | 43.9838 | 43.7375 | 0.6930 | 46.2264 | 42.6390 | 0.8535 | −18.9760 | −48.6105 | 47.07 | 47.03 |
| J2329+0020 | 0.9994 | 47.0959 | 47.0976 | 1.0000 | 43.8086 | 43.9187 | 0.9994 | 46.7381 | 41.9145 | 0.2276 | −14.3796 | −37.0532 | 47.41 | 47.16 |
| J2334−0031 | 0.6258 | 46.8182 | 46.8232 | 0.9998 | 43.8295 | 43.7356 | 0.7861 | 46.4647 | 42.4886 | 0.7864 | −17.0873 | −47.7586 | 46.91 | 46.95 |

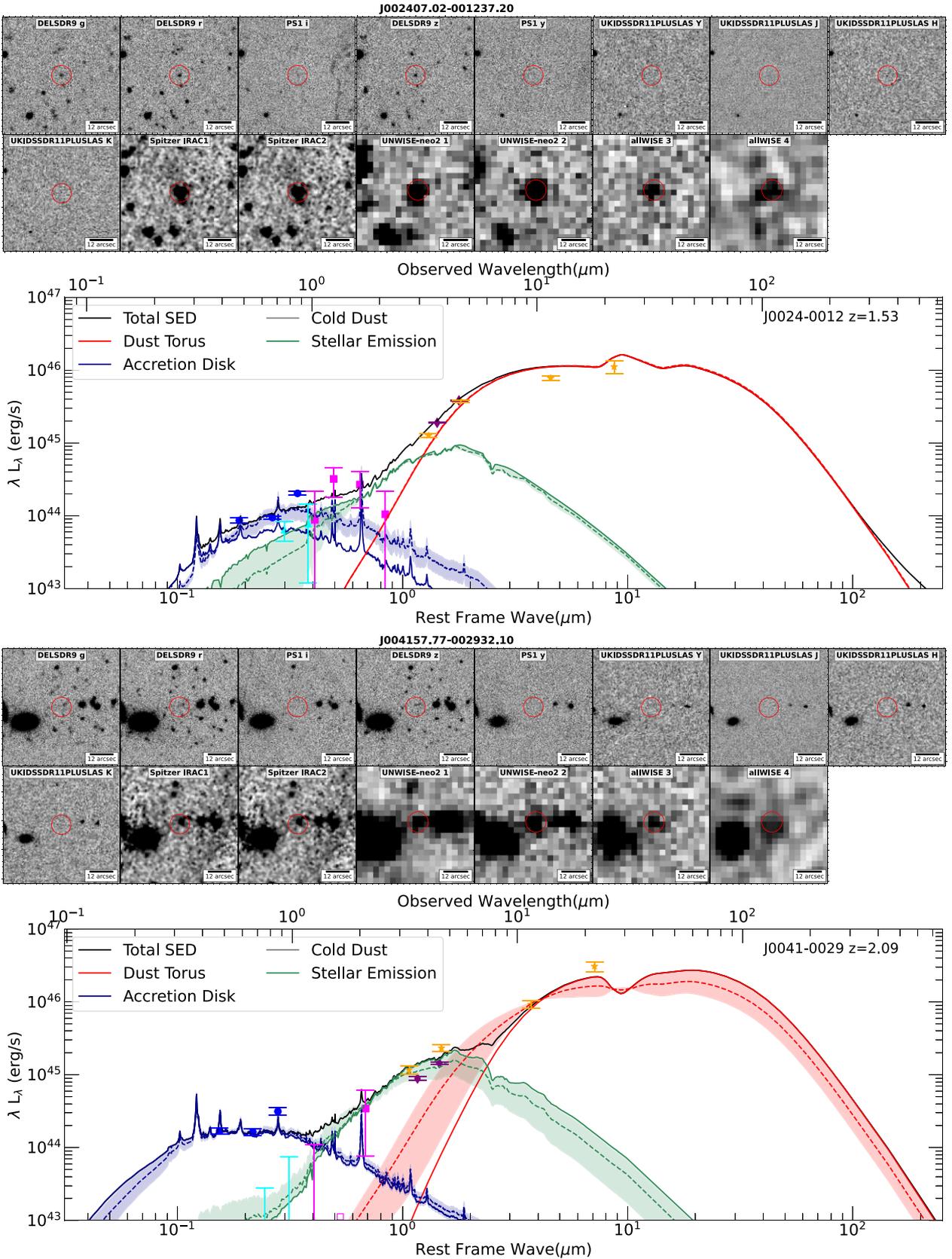

**Figure B1.** Cut-outs and SED fitting results. The SED fitting results indicate that the IR emission is dominated by the hot dust torus. We only take the dust torus as the robust fitting and use it to scale the photometry of all targets. The accretion disc and stellar emission here are not used to reveal the rest-UV/optical components. The composite rest-UV/optical emission could instead be fitted by scattered light, a reddened accretion disc, and host galaxy light; we have discussed this degeneracy in detail in Section 4.

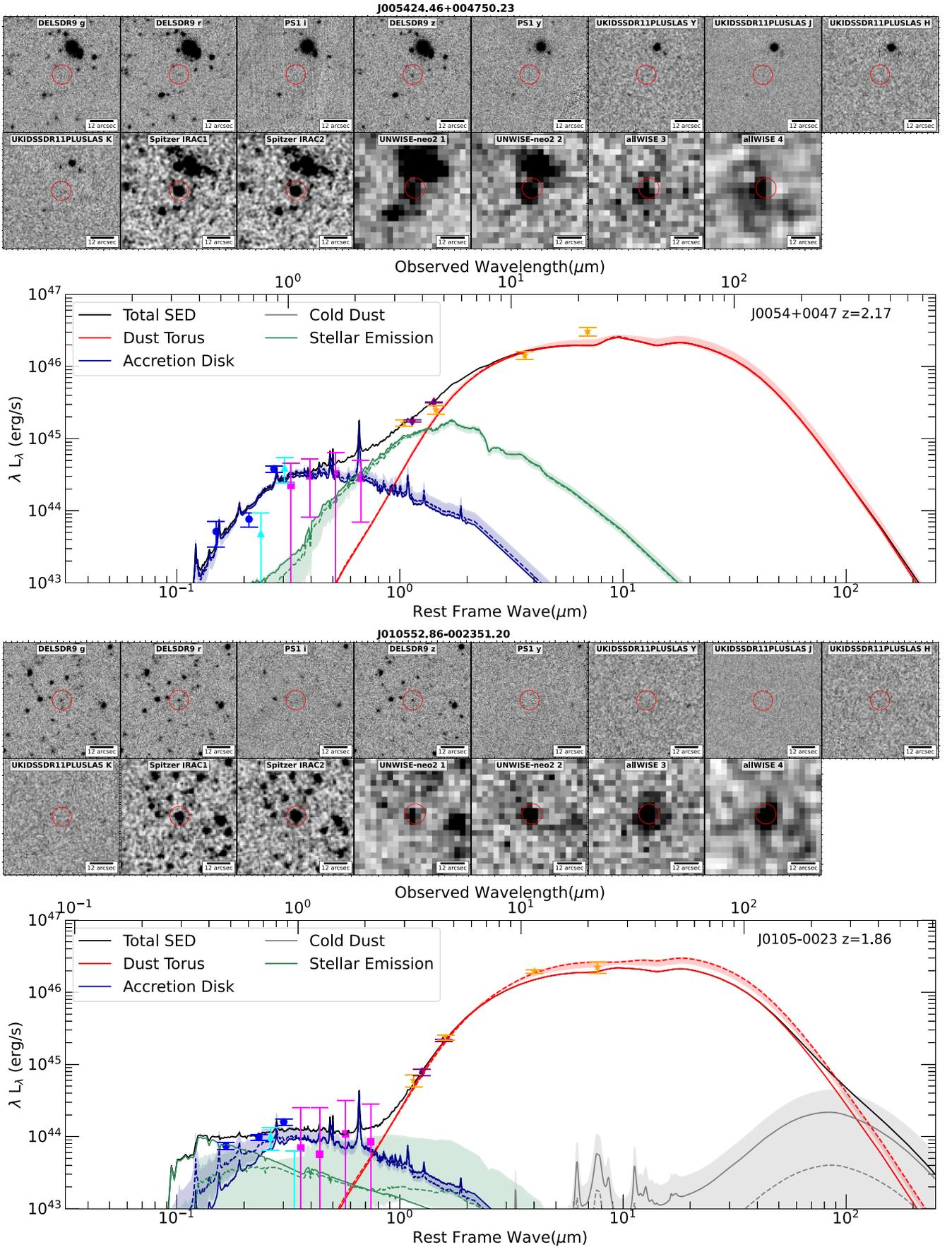

**Figure B1** — *continued*

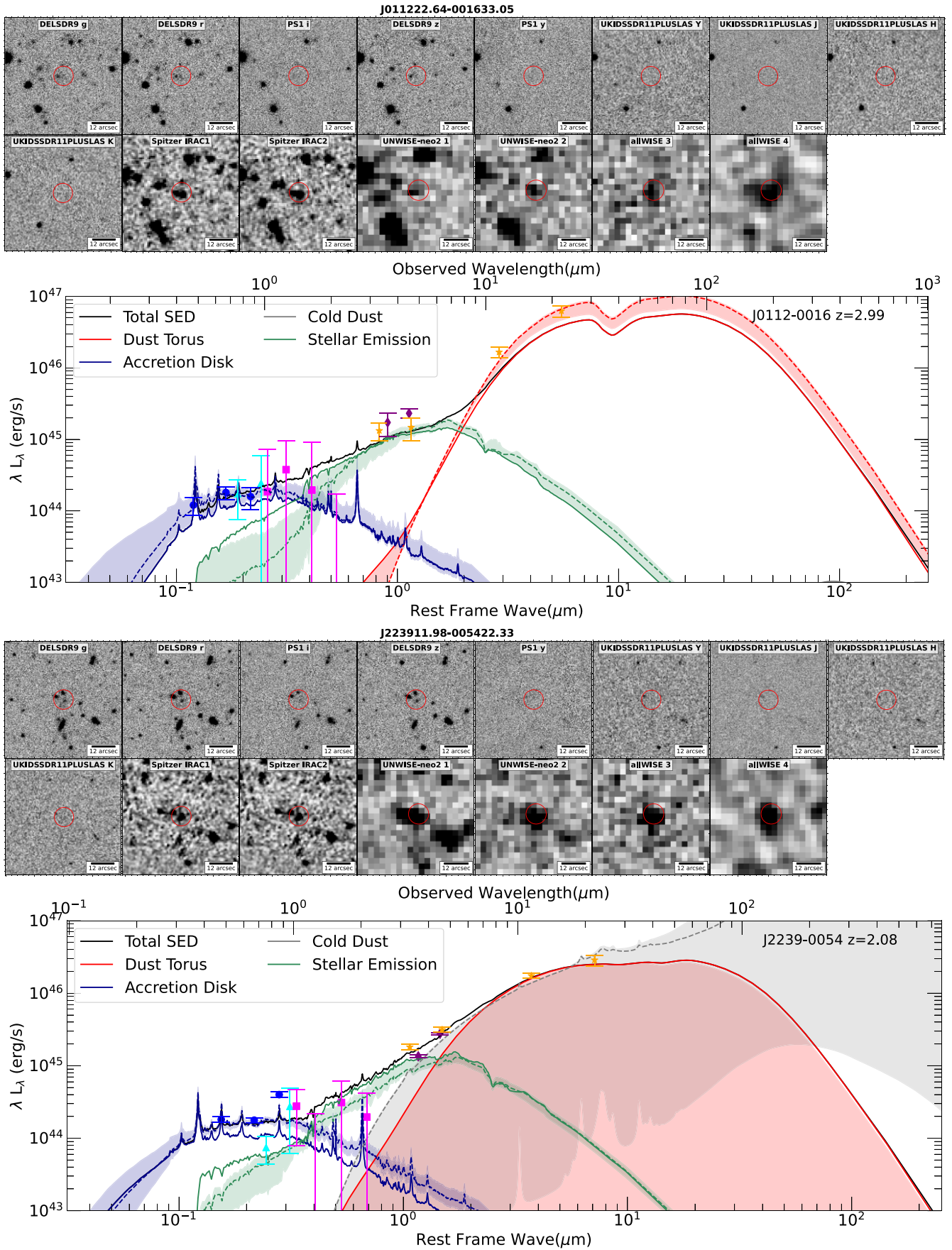

**Figure B1** – *continued*

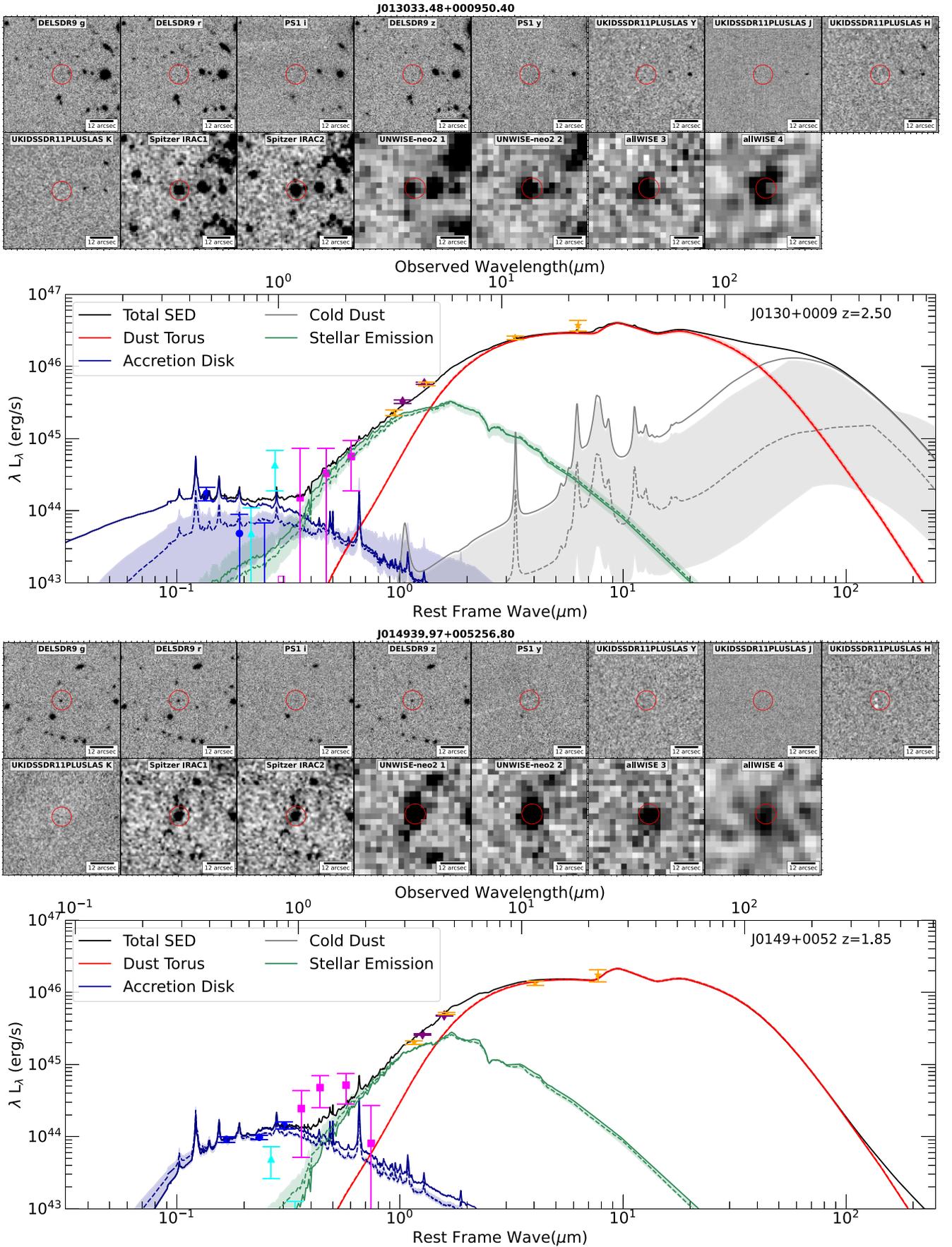

**Figure B1** – *continued*

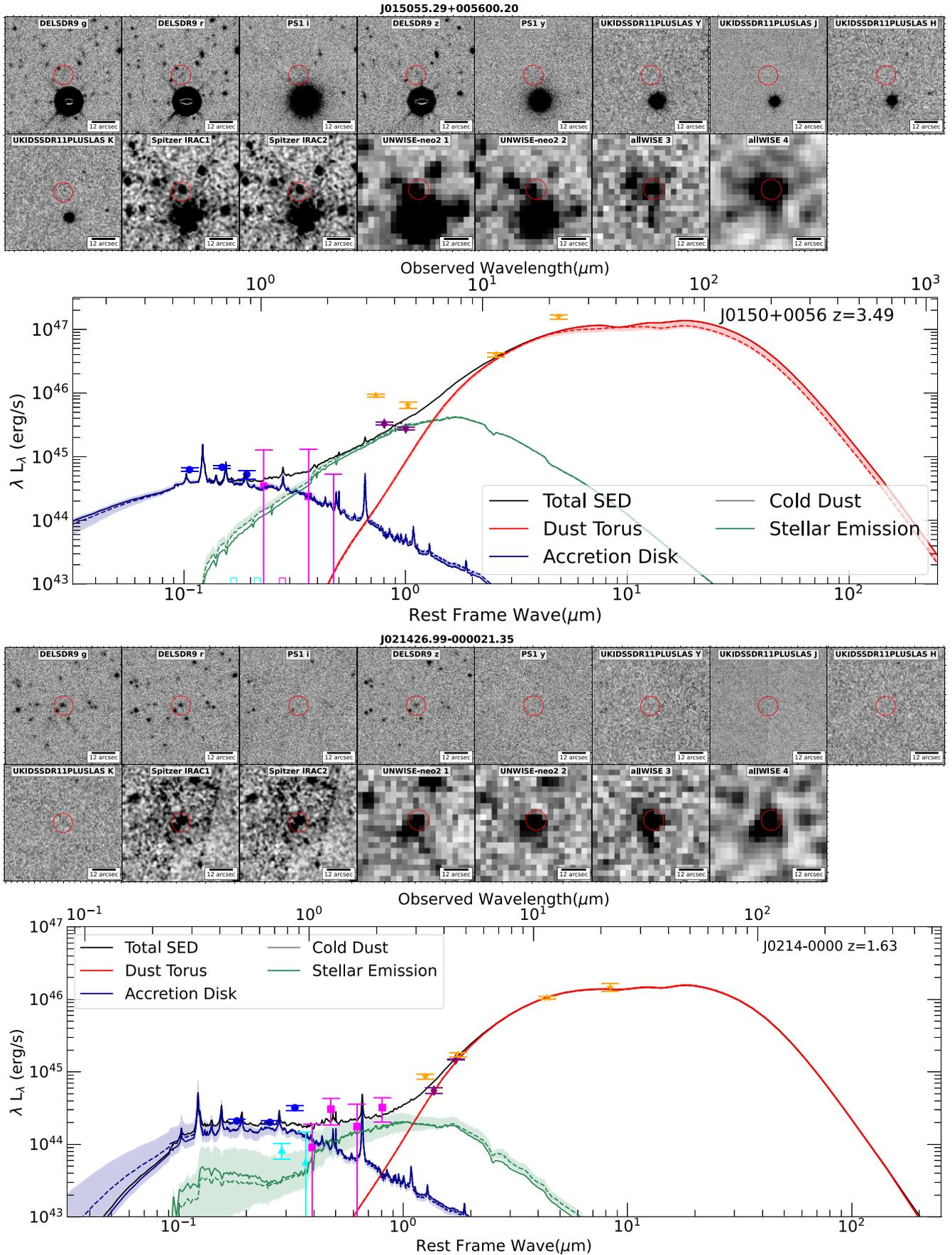

**Figure B1** – *continued*

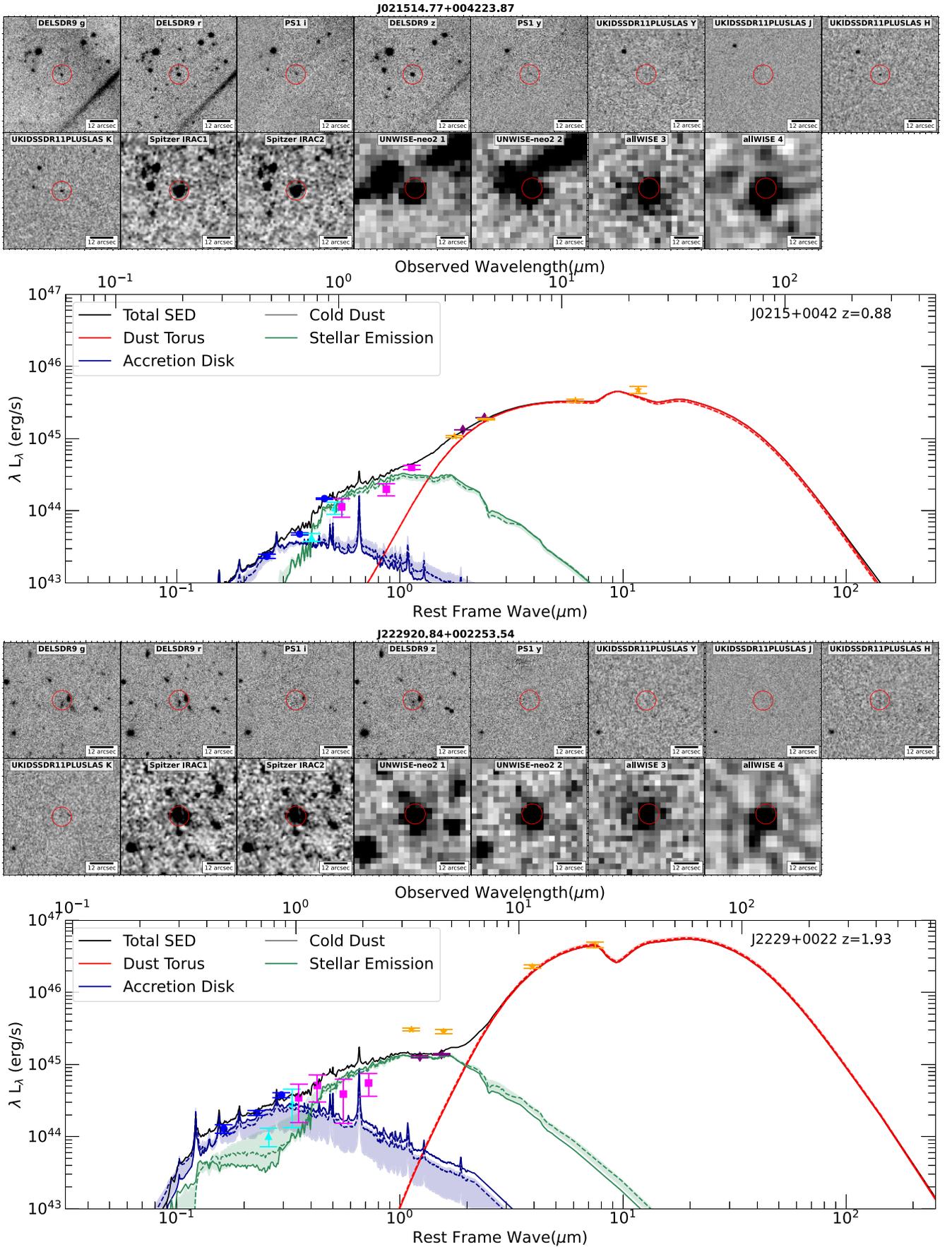

**Figure B1** – *continued*

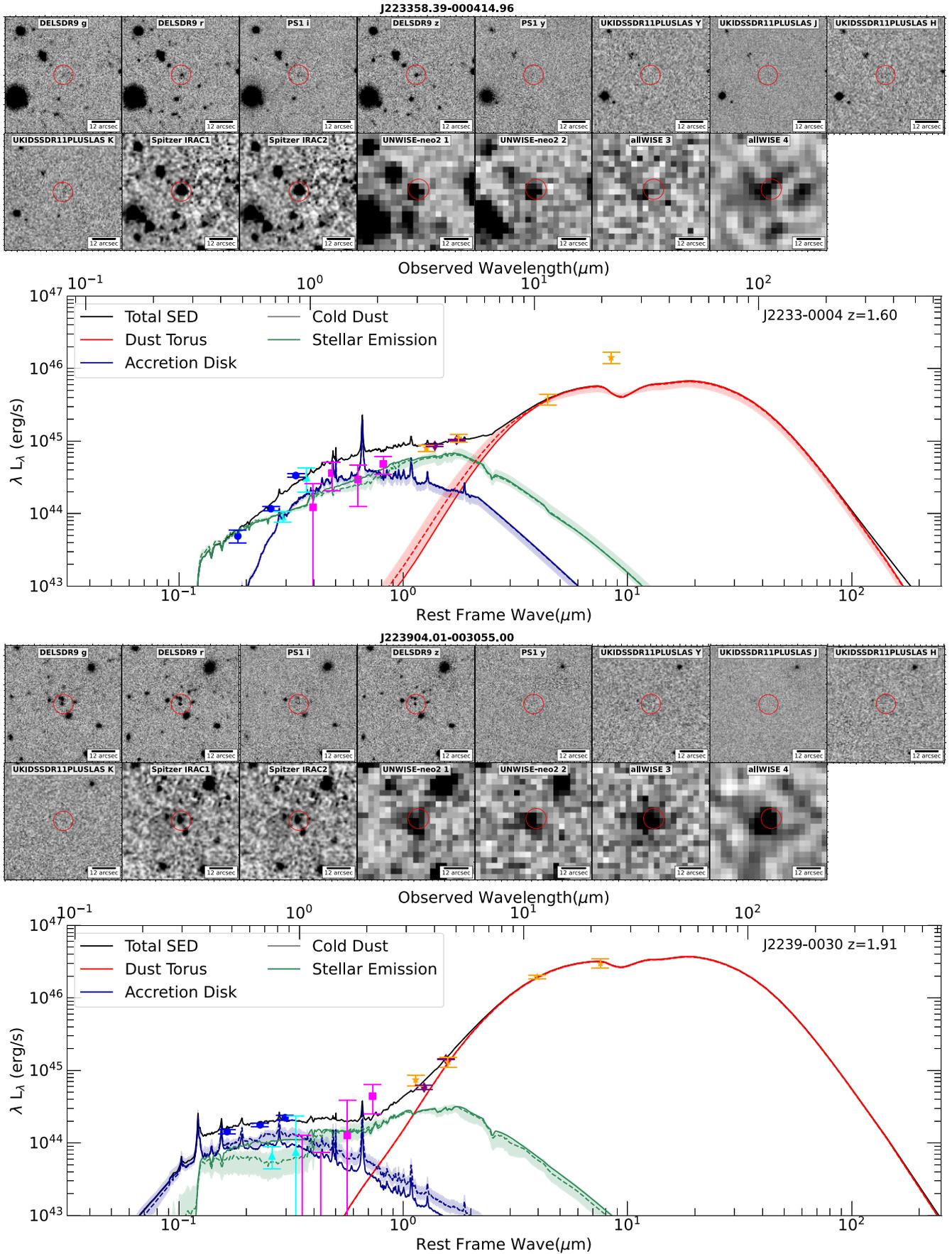

**Figure B1** *– continued*

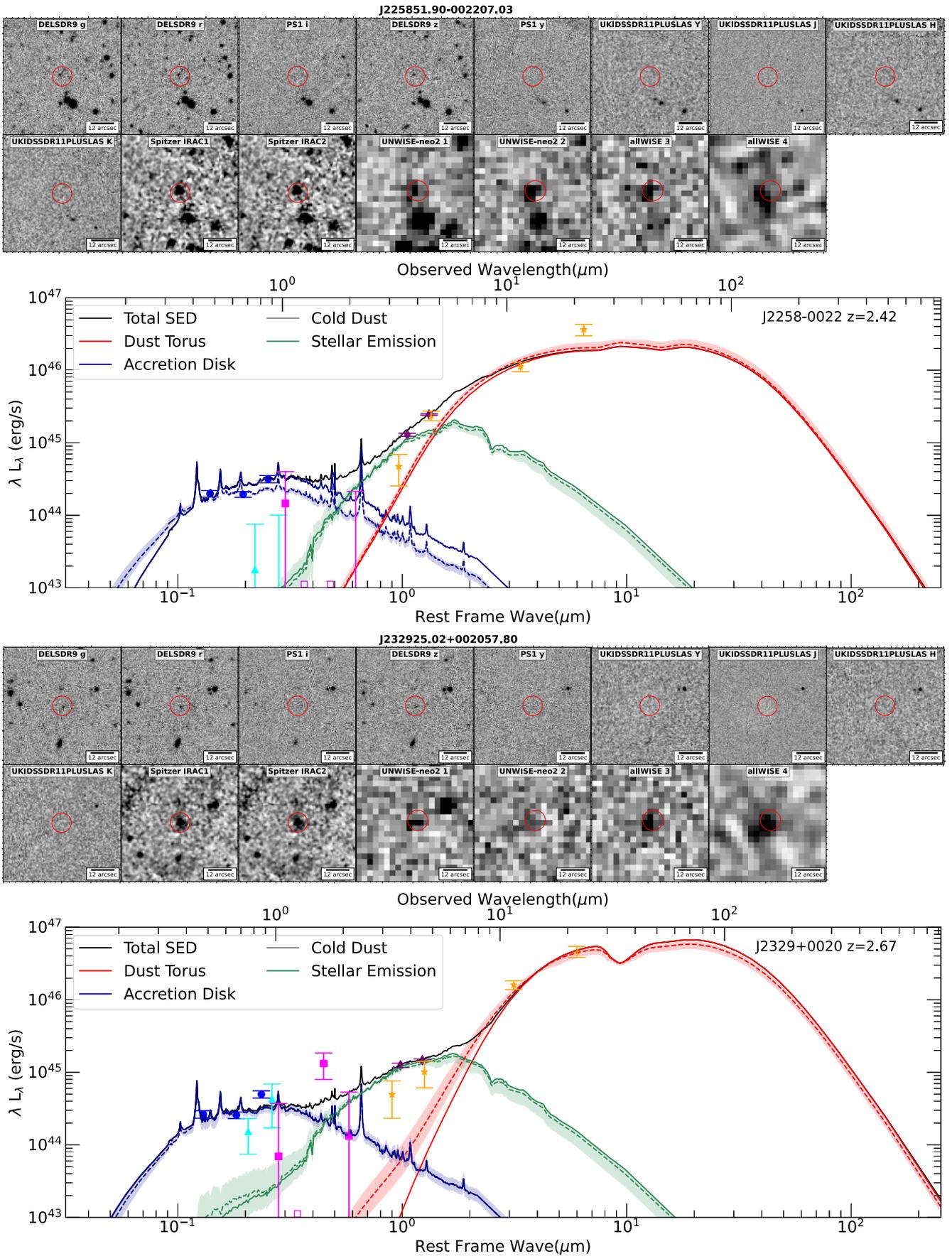

**Figure B1** *– continued*

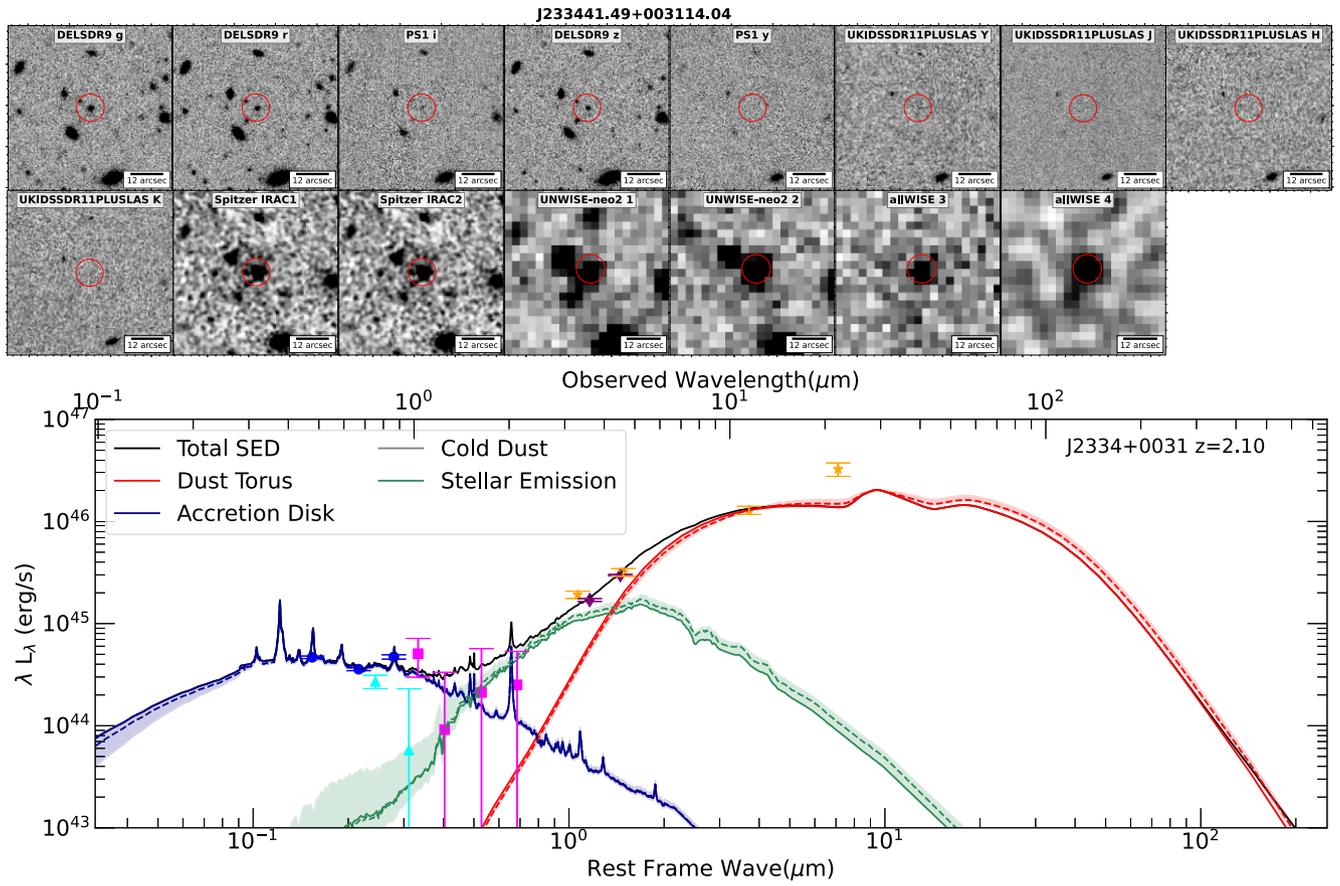

**Figure B1** – *continued*

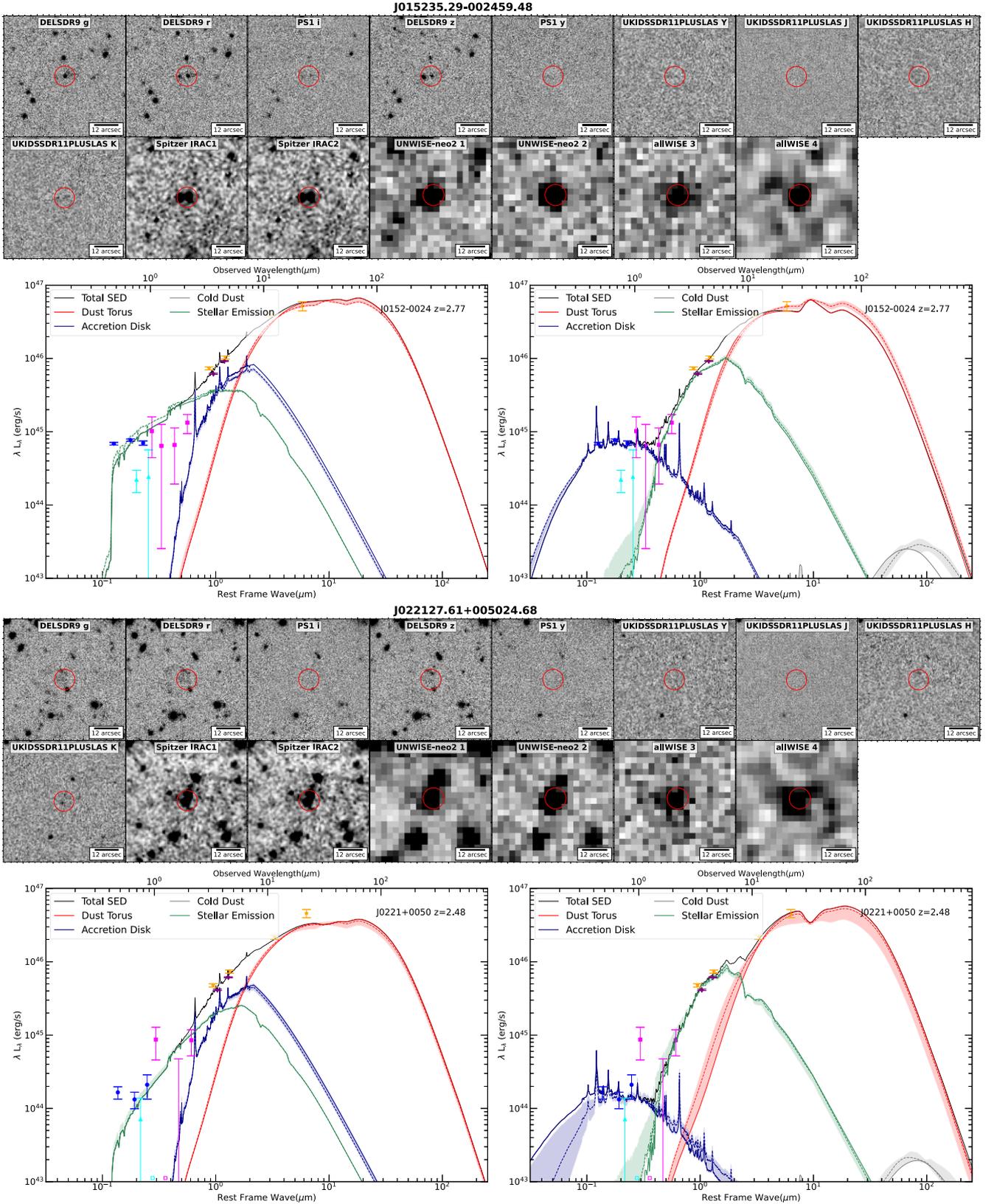

**Figure B2.** For four targets (J0152−0024, J0221+0050, J2243+0017, and J2259−0009), we plot two fitting results. If the rest-UV 1 μm light is dominated by the stellar emission, then this galaxy should have a stellar mass above $10^{12}$ M$_\odot$ to produce this brightness. This mass is higher than the typical $z \sim 2$ galaxies. Therefore, we provide another fitting result. Like in Fig. B1, only the torus is considered as the robust fitting and used to scale the photometry. The composite rest-UV/optical emission could be fitted by the scattered light, reddening accretion disc, and galaxy light; we have discussed it in detail in Section 4.

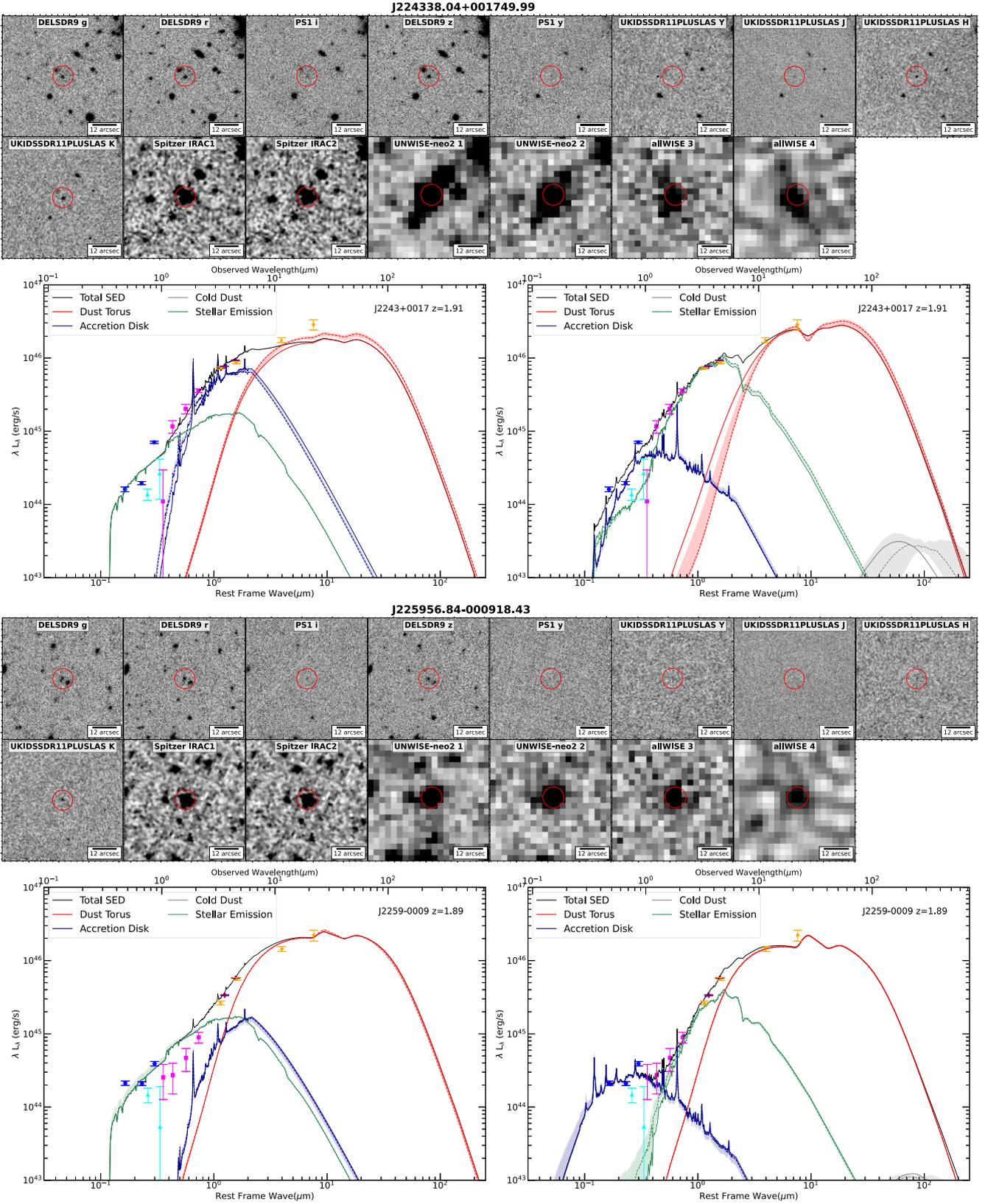

**Figure B2** – *continued*

This paper has been typeset from a T$_{\rm E}$X/L$^{\rm A}$T$_{\rm E}$X file prepared by the author.